\documentclass[reprint,superscriptaddress,amsmath,amssymb,aps,pre]{revtex4-1}

\usepackage{graphicx}
\usepackage{dcolumn}
\usepackage{bm}
\usepackage{color}
\usepackage{soul}
\usepackage{enumitem}
\usepackage{empheq}
\usepackage{flafter}     

\newcommand{\HIDDEN}[1]{}

\newcommand{\ui}{\text{i}}

\providecommand*{\oneD}{\textsc{1d}}
\providecommand*{\twoD}{\textsc{2d}}
\providecommand*{\threeD}{\textsc{3d}}
\providecommand*{\threeDD}{\textsc{3D}}
\providecommand*{\fourD}{\textsc{4d}}

\providecommand*{\someD}{\textsc{d}}

\usepackage{color}

\makeatletter
\let\Hy@backout\@gobble
\makeatother

\begin{document}

\title{Power-law trapping in the volume-preserving
       Arnold-Beltrami-Childress map}

\author{Swetamber Das}
\affiliation{Max-Planck-Institut f\"ur Physik komplexer Systeme,
             N\"othnitzer Stra\ss{}e 38, 01187 Dresden, Germany}

\author{Arnd B\"acker}
\affiliation{Technische Universit\"at Dresden, Institut f\"ur Theoretische
             Physik and Center for Dynamics, 01062 Dresden, Germany}
\affiliation{Max-Planck-Institut f\"ur Physik komplexer Systeme,
             N\"othnitzer Stra\ss{}e 38, 01187 Dresden, Germany}

\date{\today}

\begin{abstract}
  Understanding stickiness and power-law behavior of Poincar\'e recurrence
  statistics is an open problem for higher-dimensional systems, in contrast to
  the well-understood case of systems with two degrees-of-freedom.
  We study such intermittent
  behavior of chaotic orbits in three-dimensional volume-preserving systems
  using the example of the Arnold-Beltrami-Childress map.
  The map has a mixed phase space with a cylindrical regular region  surrounded
  by a chaotic sea for the considered parameters.
  We observe a characteristic overall power-law decay of the cumulative
  Poincar\'e recurrence statistics with significant oscillations superimposed.
  This slow decay is caused by orbits which spend long times close to
  the surface of the regular region.
  Representing such long-trapped orbits in frequency space shows clear
  signatures of partial barriers and reveals that coupled resonances play an
  essential role.
  Using a small number of the most relevant resonances allows
  for classifying long-trapped orbits. From this the Poincar\'e
  recurrence statistics can be divided into different exponentially decaying
  contributions which very accurately explains the overall
  power-law behavior including the oscillations.
\end{abstract}

\maketitle

\section{Introduction}
\label{sec:Intro}

Hamiltonian systems generically have a mixed phase space
in which regular and chaotic motion coexist.
A typical chaotic trajectory is
often trapped intermittently close to regular regions of phase space for
arbitrary long but finite times. The existence of such
sticky behavior is known to have important consequences for chaotic
transport, e.g.
the existence of anomalous kinetics, L\'evy processes, and L\'evy flights and
finds interesting applications in a variety of physical contexts such as
chaotic advection in
fluids~\cite{AreBlaBudCarCarCleElFeuGolGouvanKraLeMacMelMetMezMouPirSpeStuThiTuv2017},
in the three-body problem~\cite{She2010},
driven coupled Morse oscillators~\cite{SetKes2012},
confinement in particle accelerators~\cite{Pap2014}, and
plasma physics~\cite{OyaSzeBatSouCalViaSan2016},
to name a few. The behavior of
stickiness in a mixed phase-space can be characterized by the cumulative
Poincar\'e recurrence statistics $P(t)$, which is the probability that
a chaotic orbit has not returned to an initial region
until time $t$~\cite{Poi1890}.
Fully chaotic systems typically show an exponential decay of $P(t)$
\cite{BauBer1990,  ZasTip1991, HirSauVai1999, AltSilCal2004}.
However, systems with a mixed phase space usually
display a much slower decay, usually well-described by a power-law
\cite{ChaLeb1980, ChiShe1983,
  Kar1983, ChiShe1984, KayMeiPer1984a, HanCarMei1985, MeiOtt1985,
  KanKon1989, KonKan1990, DinBouOtt1990, ChiVec1993, ChiVec1997,
  ZasEdeNiy1997, BenKasWhiZas1997, ChiShe1999,
  ZasEde2000, WeiHufKet2003, AltKan2007, ShoLiKomTod2007b, CriKet2008,
  She2010, CedAga2013, AluFisMei2014, LanBaeKet2016,
  AluFisMei2017, FirLanKetBae2018}.
This so-called power-law trapping
has striking implications for transport in many physical
systems~\cite{BucDelZakManArnWal1995,BenCasMasShe2000, EzrWig2009, She2010,SetKes2012, MazShe2015}.

Power-law trapping is well-understood for Hamiltonian systems and maps
with two degrees-of-freedom ~\cite{KayMeiPer1984a,KayMeiPer1984b,MeiOtt1985,
  Mei1986,Mei1992,AluFisMei2014,AltKan2005,MotMouGreKan2005,AltMotKan2006,
  Mei2015}.
For example in area-preserving maps, regular tori are one-dimensional
and therefore act as complete barriers to transport in phase space.
If such tori break up they may turn into so-called cantori
which provide partial barriers with slow transport across them.
Usually there is a whole hierarchy of such partial barriers which
is the origin of the power-law trapping.
In contrast, in higher-dimensional systems
regular tori have an insufficient dimension to provide complete barriers to
transport. For example for a \fourD{} symplectic map,
the regular tori are two-dimensional and therefore cannot
divide the phase space into dynamically disconnected regions.
Thus already purely for topological reasons
additional types of transport are possible,
leading for example to the famous
Arnold diffusion~\cite{Arn1964,Loc1993, Dum2014}.

Still, power-law trapping is observed in higher-dimensional
systems~\cite{DinBouOtt1990,
  AltKan2007,ShoLiKomTod2007b,LanBaeKet2016,
  FirLanKetBae2018},
but not due to a generalized hierarchy
\cite{LanBaeKet2016,FirLanKetBae2018}, so that
the mechanism of stickiness is different from that in two-dimensional
area-preserving maps.
While some generalizations
have been proposed~\cite{MarDavEzr1987,Wig1990,MacMei1992,ShoLiKomTod2007b},
a profound understanding of power-law trapping in higher-dimensional
systems remains a significant open question~\cite{Mei2015}.

This motivates to investigate power-law trapping
for \threeD{} volume-preserving maps,
whether features of \twoD{} maps still apply,
while those present for \fourD{} maps might already become relevant.
For example in a \threeD{} volume-preserving map
there can be \twoD{} regular tori which therefore
are characterized by two frequencies, as in \fourD{} symplectic maps.
On the other hand, the \twoD{} tori
can provide absolute barriers to the motion
in a \threeD{} volume-preserving map
as in \twoD{} area-preserving maps.
The dynamical properties of \threeD{} volume-preserving maps
have been studied in much detail,
see e.g.~\cite{CheSun1990a, CheSun1990b, CarFeiPir1994, LomMei1998, Mez2001,
  LomMei2003, LomMei2009b, MirLom2010, Kor2010, DulMei2012, Mei2012,
  VaiMez2012, FoxMei2013, Mir2013,
  FoxLla2015, FoxMei2016, MaeSmiMit2017}.
However, there are only a few studies of power-law trapping in \threeD{}
volume-preserving maps,
see Refs.~\cite{SunZho2009, SilBeiMan2015, MeiMigSimVie2018}.
In Ref.~\cite{SunZho2009} it is argued that
stickiness in their model originates around the hyperbolic invariant sets
embedded in the sticky domain. In Ref.~\cite{SilBeiMan2015} recurrence times
statistics of an extended standard map have been analyzed.
Power-law decay in the presence of accelerator modes is found in
Ref.~\cite{MeiMigSimVie2018}.

In this paper, we investigate power-law trapping in three-dimensional
volume-preserving maps using the specific example of the
Arnold-Beltrami-Childress (ABC) map. The ABC
map is a discretized version of the ABC flows with
time-periodic forcing~\cite{DomFriGreHenMehSow1986, FeiKadPir1988}.
The ABC flows are a class of non-turbulent flows which appear
as a solution
of the three-dimensional Euler equation and is used as a simple
model for the study of chaos in three-dimensional steady incompressible
flows. The \threeD{} phase space of the flow and the corresponding map
is known to have invariant surfaces and chaotic streamlines.
The flow was first introduced by
Arnold~\cite{Arn1965} and later
Childress~\cite{Chi1970} demonstrated that these are important models for the
kinematic dynamo effect in astrophysical plasma, i.e., a magnetic field can be
generated and sustained by the motion of an electrically conducting fluid. The
flow also serves as a prototypical model of a force-free magnetic field
to study spatial diffusion of charged particles~\cite{RamDasKriMit2014}.
The ABC map exhibits most of the basic features
of a typical three-dimensional, time-periodic,
volume-preserving ABC flow. The map has a mixed phase space with periodic,
quasi-periodic, and chaotic behavior  in the \threeD{} phase space. For
specific parameters of the map,
the phenomenon of resonance-induced diffusion has
been observed which leads to global transport in phase
space~\cite{FeiKadPir1988}.
The map has been employed
as a representative of volume-preserving flows to study chaotic advection of
inertial particles in bailout embedding
models~\cite{CarMagPirTuv2002,DasGup2014,DasGup2017}.

The outline of the paper is as follows.
Section~\ref{sec:ABC_map} describes the ABC map, its basic properties,
and illustrates regular orbits in the \threeD{} phase space in
Sec.~\ref{subsec:phase_space}.
The variation of the regular region is illustrated in Sec.~\ref{subsec:FTLE}
using a sequence of \twoD{} plots of the finite-time Lyapunov exponent.
The frequency space of the system is
introduced in Sec.~\ref{subsec:Freq_space} and a boundary cylinder,
separating regular and chaotic motion, is identified.
In Sec.~\ref{sec:stickiness}, the stickiness
near the regular region is studied using the Poincar\'e recurrence
statistics and power-law trapping has been explained in Sec.~\ref{subsec:P_t}.
As illustration, one long-trapped orbit is displayed
in the \threeD{} phase space, on a \twoD{}
Poincar\'e section, and in frequency-time representation in
Sec.~\ref{subsec:long_trapped_example}.
A quantitative analysis of the approach to
the boundary cylinder is done in Sec.~\ref{subsec:approach_bc},
demonstrating the significant role played by coupled resonances.
Furthermore, in Sec.~\ref{subsec:P_t_splitting} we
use the frequency analysis to identify
a small number of the most relevant resonances.
This allows for classifying long-trapped orbits
and by this to divide the Poincar\'e
recurrence statistics into different exponentially decaying
contributions. This very accurately explains the overall
power-law behavior including the oscillations.
In Sec.~\ref{sec:partial-barriers}
possible partial barriers are briefly discussed.
Section~\ref{sec:summary} gives a  summary and outlook.

\begin{figure}[b]
	\includegraphics{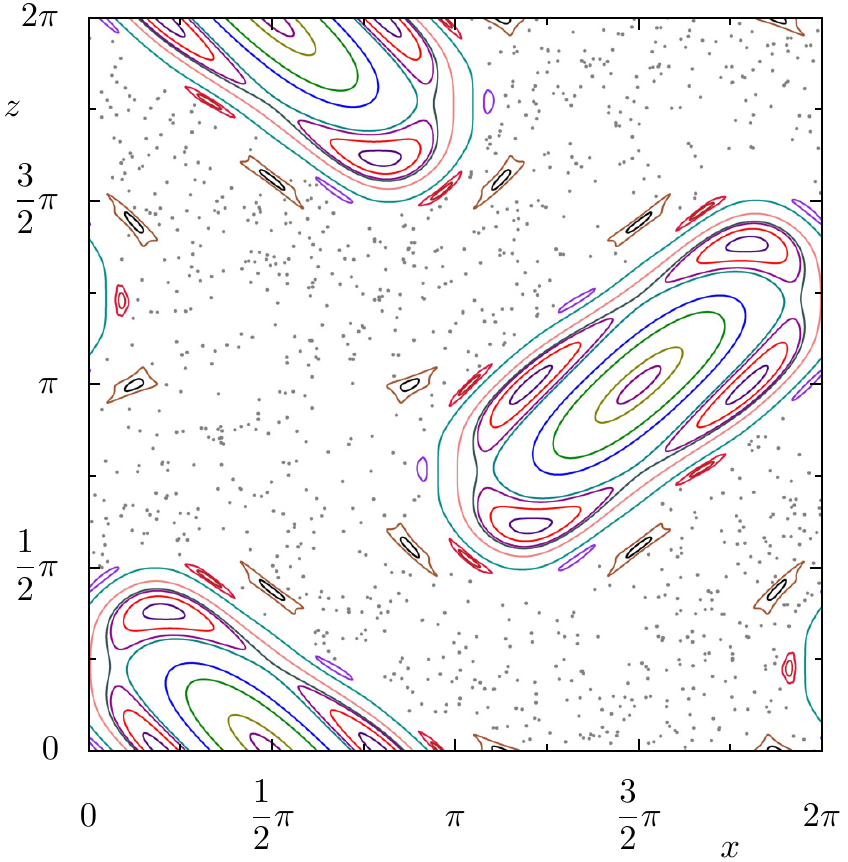}
	\caption{Phase space for the \twoD{} area-preserving map \eqref{equ:ABC_2D}
		for $(A, B) = (2.0, 1.5)$.  Shown are one chaotic orbit (gray dots) and
		16 pairs of regular tori for the central island and its symmetry
		related island in same color.
	}
	\label{fig:2D_ABCmap}
\end{figure}

\section{Arnold-Beltrami-Childress map}
\label{sec:ABC_map}

The ABC map \cite{FeiKadPir1988} is the discrete-time dynamical system
$(x_n, y_n, z_n) \mapsto (x_{n+1}, y_{n+1}, z_{n+1})$ given by
\begin{equation} \label{equ:ABC}
\begin{split}
  x_{n+1} &= x_{n} + A\sin (z_{n}) + C\cos(y_{n})  \\
  y_{n+1} &= y_{n} + B\sin(x_{n+1}) + A\cos(z_{n})  \\
  z_{n+1} &= z_{n} + C\sin(y_{n+1}) + B\cos(x_{n+1}),
\end{split}
\end{equation}
where $x_n,y_n,z_n \in [0, 2\pi[$ and periodic boundary conditions are
imposed.
The map is volume preserving for any choice of the
real parameters $A$, $B$, and $C$. The dynamics of the system
can be very different, showing a mixed phase space or
chaotic motion~\cite{FeiKadPir1988, CarFeiPir1994}.
For the rest of the paper, we fix the parameters
$(A,B) = (2.0, 1.5)$ and consider different values of $C$.

For $C=0$ the dynamics in $(x, z)$ becomes independent of the motion
in the $y$-coordinate and therefore reduces to
a \twoD{} area-preserving subsystem,
\begin{align} \label{equ:ABC_2D}
\begin{split}
  x_{n+1} & = x_{n} + A\sin (z_{n}) \\
  z_{n+1} &= z_{n} + B\cos(x_{n+1}) .
\end{split}
\end{align}
Figure~\ref{fig:2D_ABCmap} shows the \twoD{} phase space for this map
for $(A, B) = (2.0, 1.5)$,
which has the typical structure of a mixed phase
space with regular and chaotic motion.
The fixed point at $(x, z) = (3\pi/2, \pi)$ is elliptic
and therefore surrounded by regular orbits forming invariant tori
(full curves).
Embedded within this regular region one has a prominent
period-4 regular island.
In addition to this central regular island
one has a symmetry-related island around the elliptic
fixed point at $(x, z) = (\pi/2, 0)$.
Outside of these regular islands one has a large area
with predominantly chaotic motion,
as illustrated by a longer orbit in the figure.
Due to the symmetry, dynamical properties, such as stickiness
at the islands are identical, so that we
can restrict to the central region around the
fixed point at $(x, z) = (3\pi/2, \pi)$.

\subsection{\threeDD{} phase space}
\label{subsec:phase_space}

\begin{figure}
\includegraphics{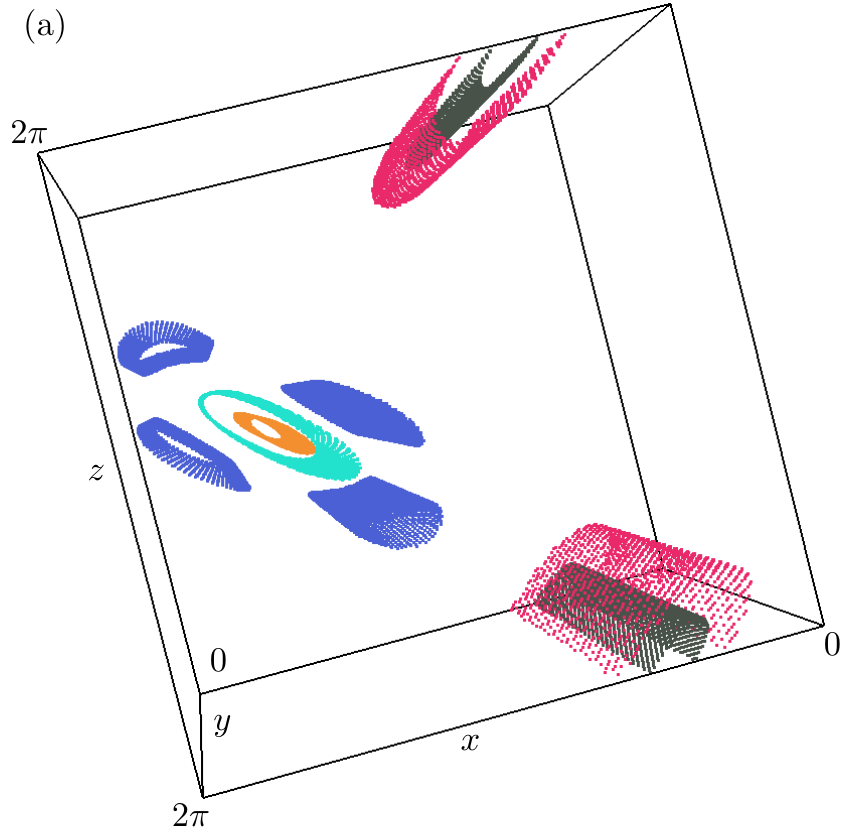}\\
\vspace*{1cm}
\includegraphics{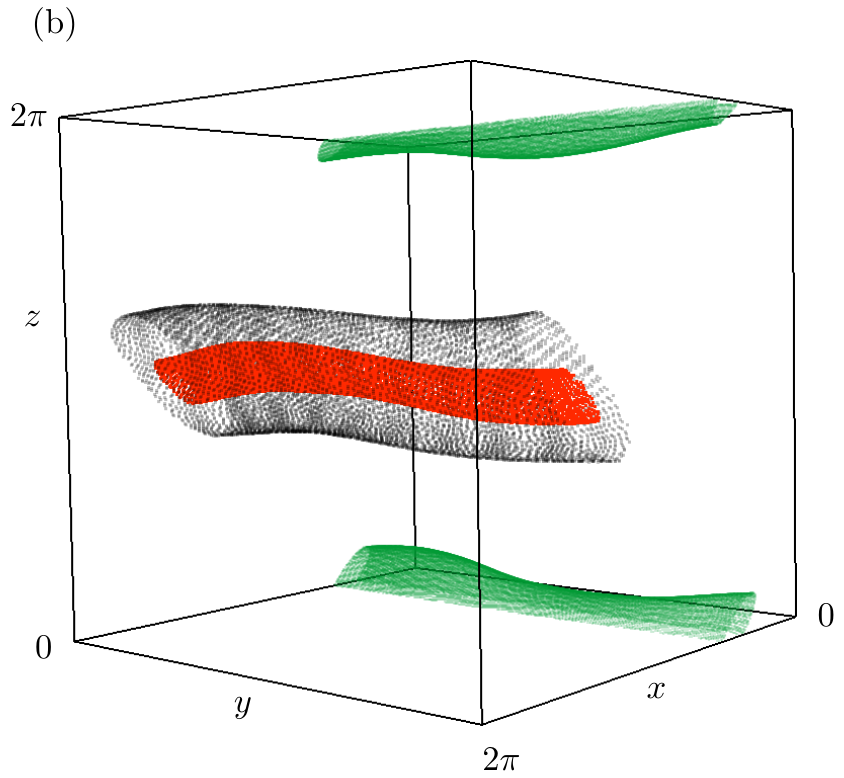}

  \caption{Plots of \twoD{} tori in the \threeD{} phase space
    of the ABC map \eqref{equ:ABC}
    for $(A, B) = (2.0, 1.5)$ and (a) $C=0$ and (b) $C=0.06$.
    }
  \label{fig:ABC3D}
\end{figure}

When considering the full \threeD{} map \eqref{equ:ABC}
for $C=0$, one has in addition
to the \twoD{} dynamics \eqref{equ:ABC_2D}
the motion in $y$-direction given by
\begin{equation}
  y_{n+1} = y_{n}+B\sin(x_{n+1}) +A\cos(z_{n}).
\end{equation}
As the \twoD{} system \eqref{equ:ABC_2D} decouples from the
motion in $y$-direction when $C = 0$,
the stable fixed points and periodic points
of the \twoD{} map turn into elliptic
\oneD{} invariant lines for the \threeD{} map.
The surrounding regular tori turn
into \twoD{} regular tori forming concentrically nested cylinders.
This is illustrated in Fig.~\ref{fig:ABC3D}(a).
Unstable fixed points and unstable periodic points
of the \twoD{} map turn into hyperbolic \oneD{} invariant lines.

\begin{figure}
\includegraphics{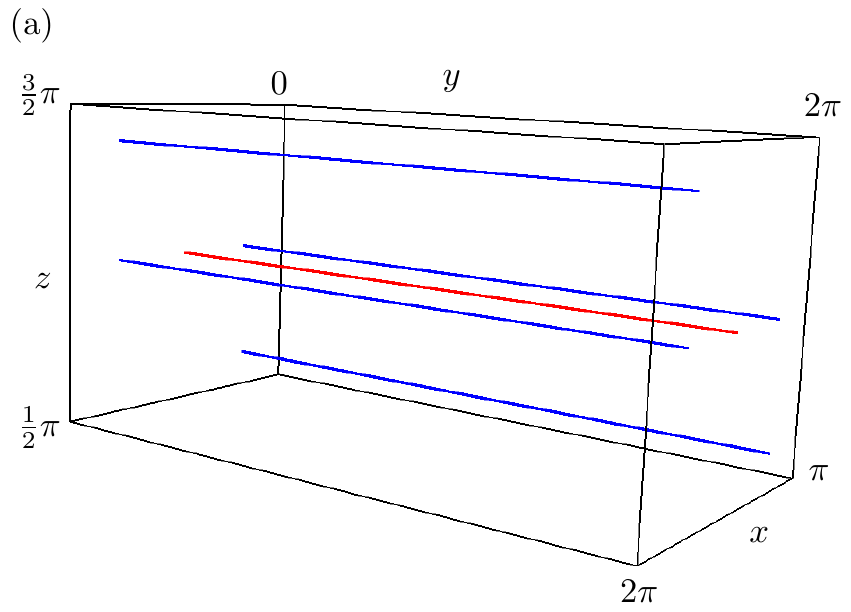}\\
\includegraphics{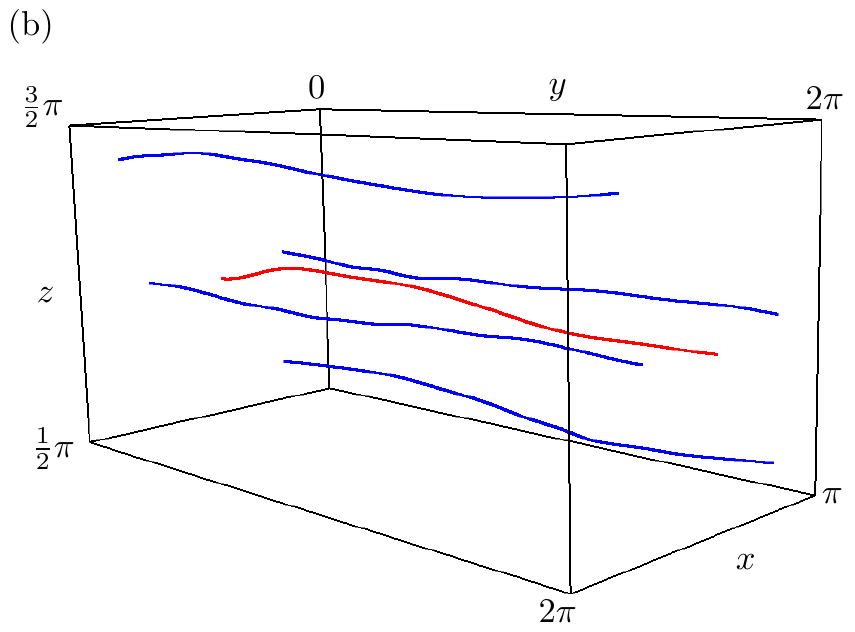}
  \caption{Location of \oneD{} tori for (a)
    $C = 0.0$ and (b) $C = 0.06$ of the
    central regular region.
    (a) For $C = 0.0$ there are four \oneD{} tori corresponding
    to the periodic orbit of period 4 of the \twoD{} map \eqref{equ:ABC_2D}
    and one central \oneD{} torus.
    (b) For $C = 0.06$  all these \oneD{} tori are deformed.
    }
  \label{fig:one_d_tori_C_6}
\end{figure}

For $C>0$ the dynamics in the $(x, z)$ subsystem couples
with the motion in $y$-direction.
Starting from $C=0$ and increasing $C$, the lines of the \oneD{}
tori begin to distort gradually and the surrounding cylinders of \twoD{} tori
also deform.
This is illustrated in Fig.~\ref{fig:ABC3D}(b)
for the \twoD{} tori and in Fig.~\ref{fig:one_d_tori_C_6} where
five \oneD{} tori are shown,
which for $C=0$ form straight lines in the \threeD{} phase space
and become distorted with increasing $C$.
We obtain these \oneD{} tori by the ``contraction method"
starting from \twoD{} tori and iteratively
reducing the radius in the $(x, z)$ plane \cite{LanRicOnkBaeKet2014}.

Such elliptic \oneD{} tori are surrounded by \twoD{} tori.
If these form a cylinder extending along the whole $y$-direction,
they act as a full barrier to the motion, i.e.\
orbits started inside will never be able
to move outside, and vice versa.
This is therefore similar to the case of \twoD{} area-preserving
maps, where \oneD{} tori form absolute barriers
to the motion \cite{Mei1992, Mei2015}.
With increasing perturbation, more and more of these \twoD{} tori break up
and resonances can lead to gaps in the families of \twoD{} tori,
see Sec.~\ref{subsec:Freq_space} below.

\begin{figure}

   \includegraphics{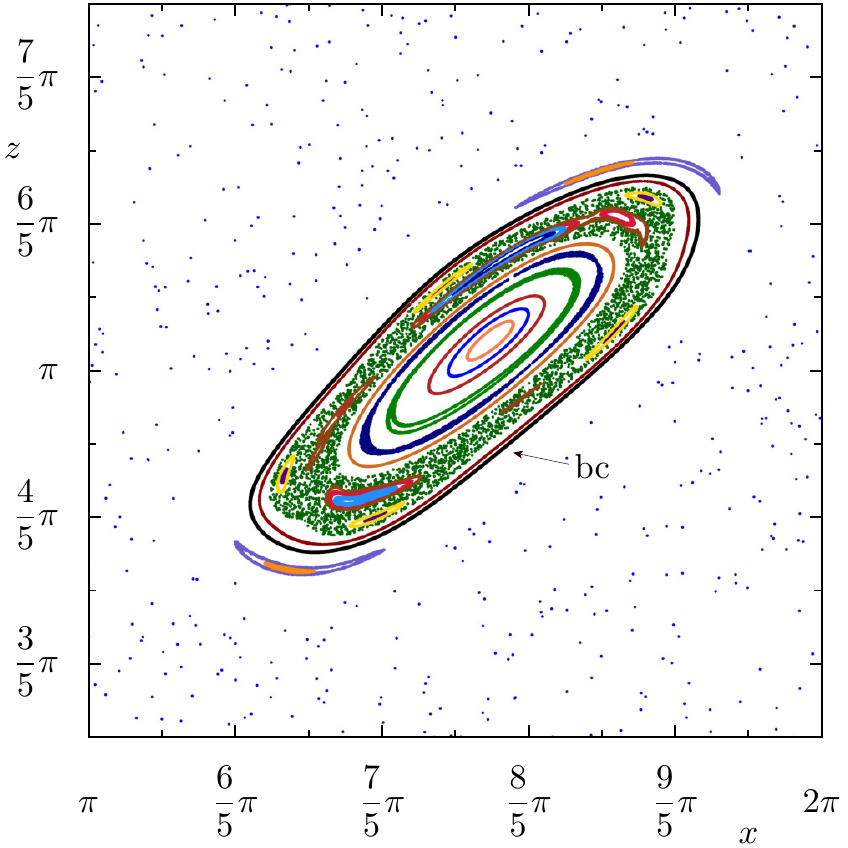}
  \caption{Plot of orbits of the \threeD{} ABC map
    in the \twoD{} slice \eqref{eq:slice-condition}.
    The blue dots represent a chaotic orbit.
    The orbits corresponding to the \twoD{} tori shown
    in \ref{fig:ABC3D}(b)
     belong to this central island. There exist chaotic layers inside the
     regular island. One such layer is shown in green. The
     outmost curve in black represents the boundary cylinder (bc)
     which separates the inside of the ``regular'' region from
     the chaotic sea.
   }
  \label{fig:2D-slice}
\end{figure}

A representation similar to Fig.~\ref{fig:2D_ABCmap}
can be obtained for the \threeD{} map by using a \twoD{} slice at $y=\pi$:
only those points $(x_n, z_n)$ of an orbit for which
\begin{equation} \label{eq:slice-condition}
  |y_n - \pi| < \varepsilon = 10^{-3}
\end{equation}
holds, are plotted, see Fig.~\ref{fig:2D-slice} for $C=0.06$.
The parameter $\varepsilon$ determines the resolution of the resulting plot
and smaller values require more iterations to obtain
the same number of points of an orbit in the slice.
While there are orbits which never fulfill the
slice condition \eqref{eq:slice-condition},
a lot of the geometry of the regular orbits
can be seen in this plot.
Usually \oneD{} tori correspond to single points in the \twoD{} slice
and \twoD{} tori lead to one or several loops.
Thus the visual appearance is similar to that of the \twoD{}
map shown in Fig.~\ref{fig:2D_ABCmap}.
In particular there still seems to be some
outer regular curve which corresponds to a \twoD{}
cylinder in the \threeD{} phase space.
Numerically no apparent holes in this surface have been found.
Thus such an outer \twoD{} torus forms an absolute barrier
separating the regular region from the chaotic sea.
Therefore it can be considered
as an analogue of the boundary circle in \twoD{} maps \cite{GreMacSta1986}.
We will refer to this as boundary cylinder in the following,
see Sec.~\ref{sec:stickiness} for further discussion on its numerical
determination.

Additionally, inside the region enclosed
by the boundary cylinder there are some chaotic regions,
of which the most prominent one is shown as green dots.
It seems that none of these inner chaotic regions is dynamically connected
to the outer chaotic sea. Several smaller regular islands are
embedded within the chaotic region, some of them
are shown in Fig.~\ref{fig:2D-slice}.
The orbit in blue is a chaotic orbit which stays outside of the region enclosed
by the boundary cylinder.

\begin{figure*}
  \includegraphics[width=1.0\textwidth]{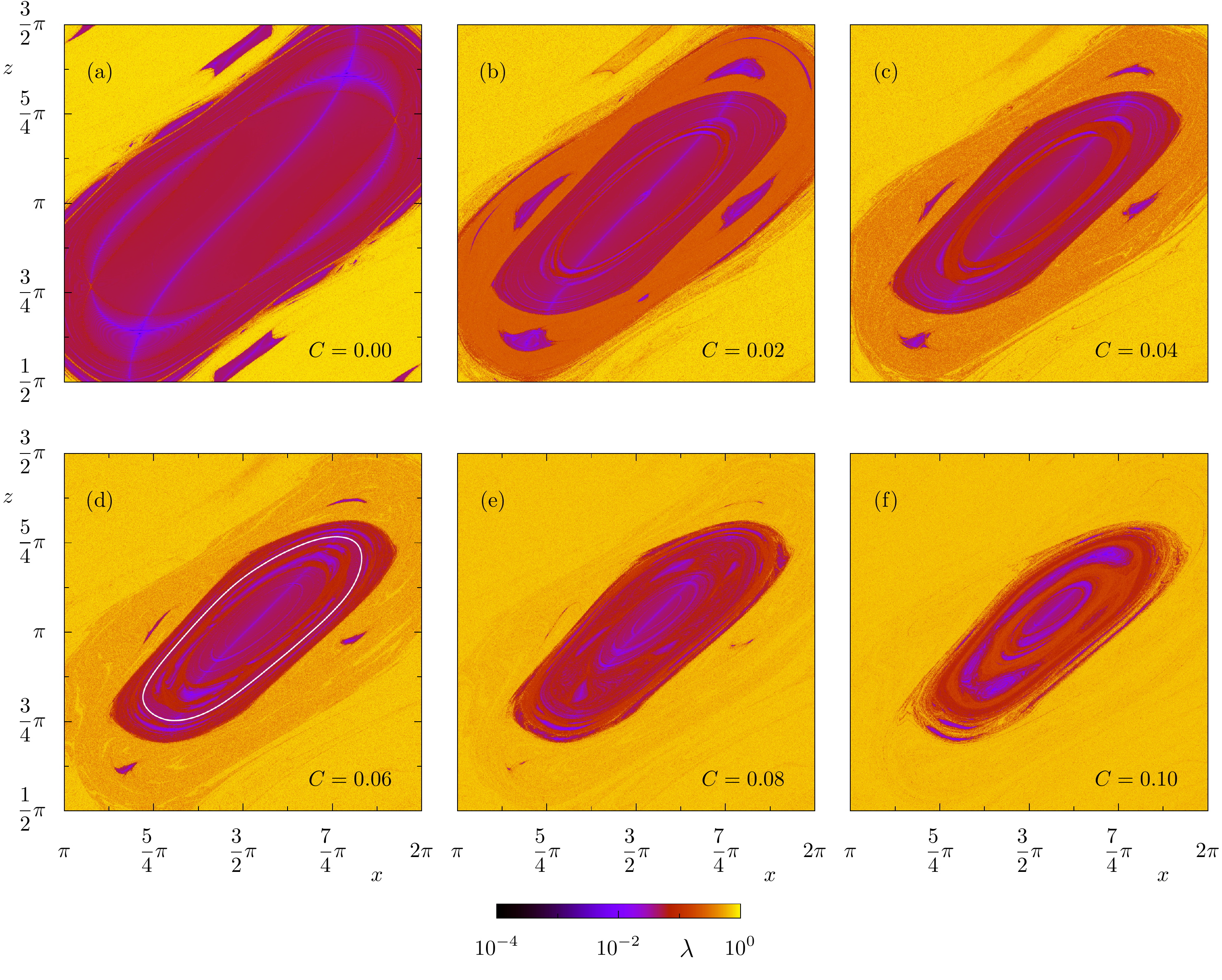}

  \caption{Finite-time Lyapunov exponent $\lambda$ for the ABC map around the
    central regular region on the $y = \pi$ plane for
    (a) $C =
    0.0$, (b) $C = 0.02$, (c) $C =
    0.04$, (d) $C = 0.06$, (e) $C = 0.08$, and (f) $C = 0.10$.
    In (d) the boundary cylinder, restricted to the \twoD{} slice, is
    shown as a thin white line.}
    \label{fig:Lyap_exp_plot}

  \vspace{1cm}
  \includegraphics[width=1.0\textwidth]{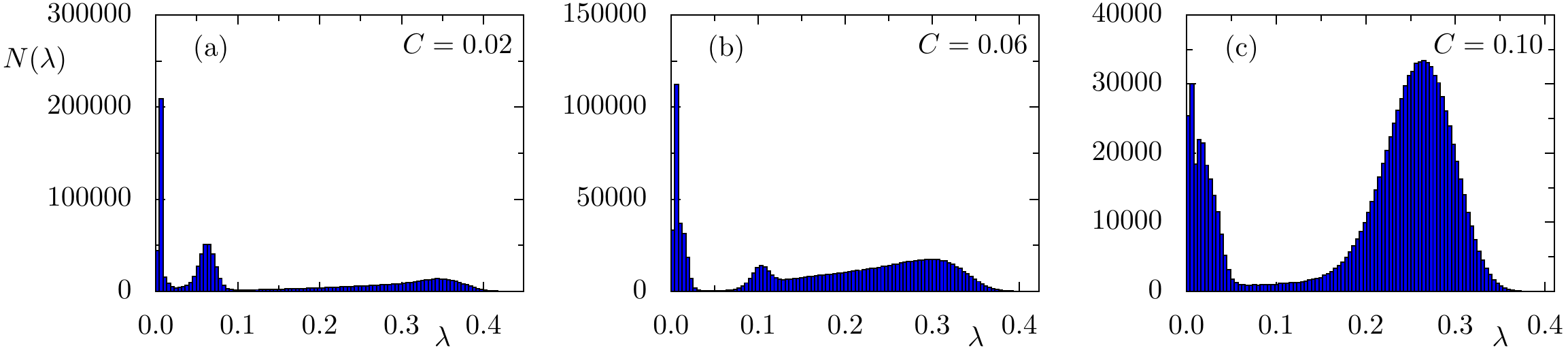}
\caption{Distribution $N(\lambda)$
    of the finite-time Lyapunov exponents $\lambda$
    for the ABC map for the plots shown in Fig.~\ref{fig:Lyap_exp_plot}
    for (a) $C=0.02$, (b) $C=0.06$, and (c) $C=0.1$. An initially
    tri-modal distribution in (a) gradually changes to bimodal in (c).
}
  \label{fig:Lyap_hist}
\end{figure*}

\subsection{Finite-time Lyapunov exponent}
\label{subsec:FTLE}

To get an overview of how the regular structures are
affected when increasing the coupling $C$ we
use the finite-time-Lyapunov exponent (FTLE) as
chaos indicator~\cite{EckRue1985,SzeLopVia2005,ManBeiRos2012}.
The FTLE estimates the rate of separation of orbits started at nearby
initial conditions. Large values of the FTLE indicate chaotic motion.
Figure~\ref{fig:Lyap_exp_plot} shows the FTLE for a sequence of values for $C$,
computed in the plane $y=\pi$
and a grid of $1000 \times 1000$ points for
$x \in\left[\pi,2\pi\right] $ and $z \in \left[\pi/2,3\pi/2\right]$,
corresponding to the surrounding of the central regular region.
For $C= 0$, regular motion with small FTLE  appears as a large island.
Already for $C=0.02$ the region with small FTLE has substantially
reduced and a significant outer part of the former regular region
has turned into a region with intermediate values of the FTLE.
Typically, initial conditions in this region will stay there
for long time before they eventually leave towards the chaotic sea.
This sticky region is of particular relevance
for the power-law trapping studied in Sec.~\ref{sec:stickiness}.
Embedded in this region is the period-4 elliptic \oneD{} torus
and its surrounding \twoD{} tori, as reflected in the smaller FTLE.

With increasing $C$, the values of the intermediate FTLE
become larger until at $C=0.1$ the same values
as in the surrounding chaotic sea are obtained.
Thus the regular region erodes from the outside with increasing coupling
$C$. At the same time also the regular orbits around
the former period-4 elliptic \oneD{} tori
are no longer visible.
Finally, at $C = 0.1$, see Fig.~\ref{fig:Lyap_exp_plot}(f),
the main regular island shrinks considerably.

This gradual transition is also reflected in the sequence
of histograms of the FTLE shown in Fig.~\ref{fig:Lyap_hist}.
For $C=0.02$ one has a large peak at small values of the FTLE
indicating the large regular region.
Chaotic motion corresponds to the peak at larger values
and the peak in between corresponds to the intermediate
type of motion surrounding the central regular part seen in
Fig.~\ref{fig:Lyap_exp_plot}(b).
With increasing $C$ this peak moves towards larger values of the FTLE,
indicating more chaotic motion, and eventually merges at $C=0.1$.
with the peak associated with the chaotic sea.
In addition, the fraction of regular orbits substantially
decreases with increasing $C$.

Interestingly, the way how the \twoD{} tori break up
for the considered ABC map leads to
a quite large chaotic layer inside the regular island, see
Fig.~\ref{fig:2D-slice}, which is responsible
for the additional peak in the FTLE histograms.
This seems to be more pronounced than
that for \oneD{} tori in a typical area-preserving map.
For example the FTLE distribution for the standard map makes a transition
from bimodal to unimodal (Gaussian) when increasing the chaoticity
in the system~\cite{SzeLopVia2005}. Also for other parameters
the histograms appear essentially bimodal.

\subsection{Frequency space}
\label{subsec:Freq_space}

\begin{figure*}[t]
  \includegraphics{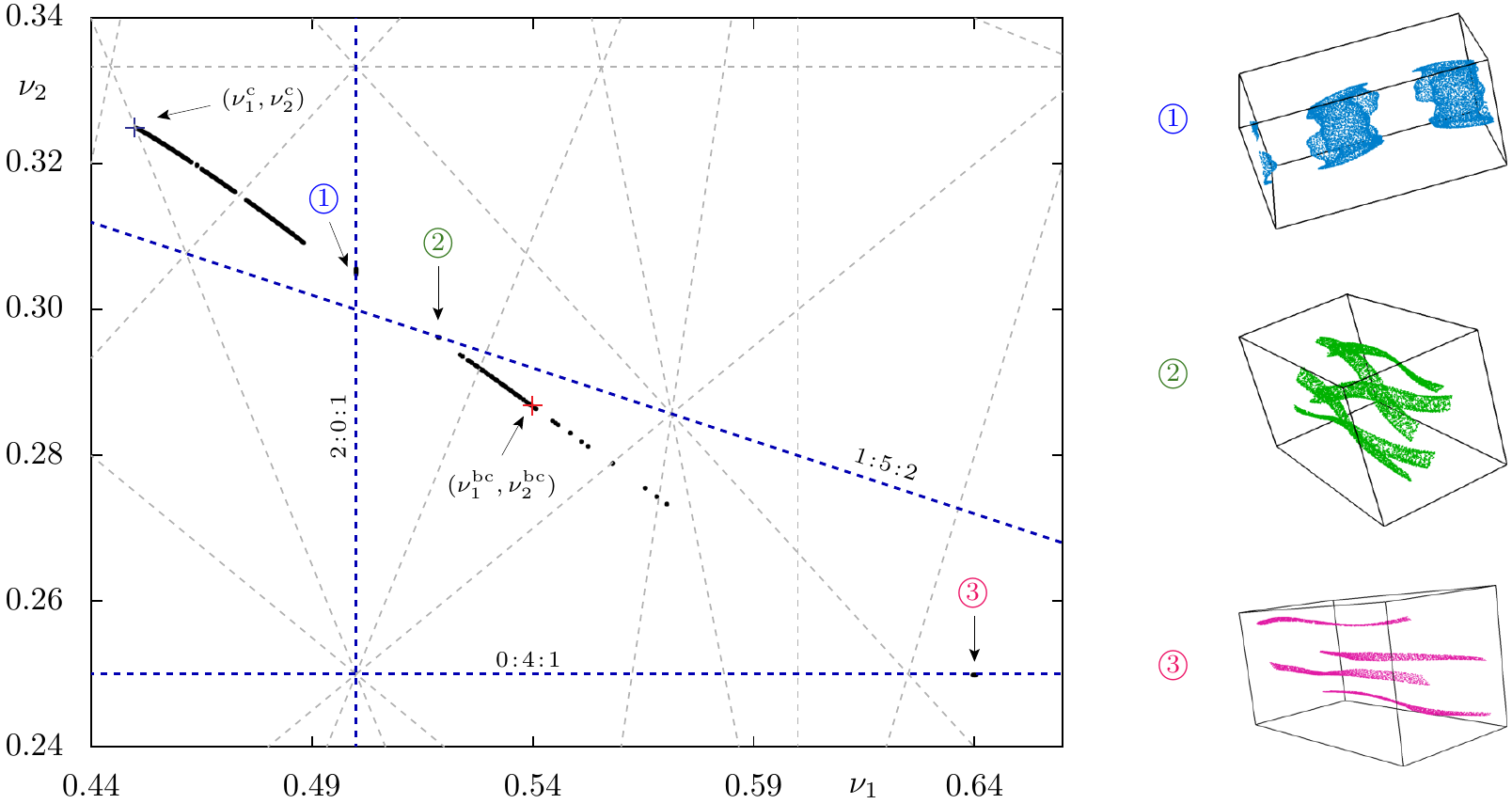}

  \caption{Frequency space of the ABC map~\eqref{equ:ABC}
    for $(A,B,C) = (2.0, 1.5, 0.06)$ showing
    $2.3\times10^6$ frequency pairs (black dots),
    each corresponding to a regular \twoD{} torus
    of the central regular region.
    Some selected resonance lines are shown as dashed lines.
    For three resonances corresponding orbits are
    shown to the right: (1) $2:0:1$ (blue),
    (2) $1:5:2$ (green), and (3)  $0:4:1$ (pink).
    The location of the boundary cylinder
    $(\nu^\text{bc}_1, \nu^\text{bc}_2)$ is indicated as
    red cross and the  central \oneD{} torus
    $(\nu^\text{c}_1, \nu^\text{c}_2)$ as blue cross.}
  \label{fig:freq_reg_tori}
\end{figure*}

The frequency-space representation is an important complementary approach to
investigate the organization of regular motion
in higher-dimensional dynamical systems.
With every regular \twoD{} torus in the \threeD{} phase space
we associate two fundamental frequencies $(\nu_1,\nu_2) \in [0, 1)^2$
and display them in a two-dimensional frequency space.
This is numerically done using a Fourier-transform-based
frequency analysis~\cite{MarDavEzr1987,Las1993,BarBazGioScaTod1996}.
For a given initial condition $(x_0, y_0, z_0)$
we use 4096 points of the orbit and compute
the fundamental frequencies $\nu_1$ and $\nu_2$
using two complex signals $\cos y_n + \ui \sin y_n$ and
$x_n + \ui z_n$ respectively.
These correspond to the motion in $y$-direction and the projection
on the $(x, z)$-plane, respectively.

To decide whether an orbit is regular or chaotic,
two consecutive segments of 4096 points are compared,
giving frequencies   $(\nu_1,\nu_2)$ and $(\tilde{\nu}_1,\tilde{\nu}_2)$.
As chaos indicator we use
\begin{equation} \label{eq:freq-delta}
\delta = \text{max}(|\nu_1 - \tilde{\nu}_1|,|\nu_2 - \tilde{\nu}_2|)
       < \delta_\text{reg}.
\end{equation}
For a regular orbit, the frequencies of the consecutive segments
are very similar so that $\delta$ is very small.
On the other hand, for a chaotic orbit, the frequencies of successive segments
will be very different.
Thus an orbit is accepted as regular if $\delta < \delta_{\text{reg}}$.
It is important to point out that this analysis is based
on finite-time information.
Hence, some orbits which are accepted as regular \twoD{} tori may
actually be chaotic orbits. However, this is a
common limitation of any tool for chaos detection,
see e.g.~Ref.~\cite{SkoGotLas2016}.

For the central regular region we use $10^8$ random initial conditions with
$x \in\left[\pi,2\pi\right]$, $z \in \left[\pi/2,3\pi/2\right]$
and $y \in[0, 2\pi[$ and use $\delta_{\text{reg}} = 10^{-9}$.
This gives a total of  $2.3\times10^6$
regular tori for $C=0.06$. These frequency pairs are shown in the
two-dimensional frequency space in Fig.~\ref{fig:freq_reg_tori}.
They are essentially arranged along a one-dimensional curve,
emanating from the point of the frequencies of the central \oneD{} torus
at $(\nu_1^{\text{c}}, \nu_2^{\text{c}}) \approx (0.4495 ,0.325)$.
This illustrates that the regular orbits
form a one-parameter family in the volume-preserving case
\cite{DulMei2009,DulMei2012}.
Note that in contrast in \fourD{} symplectic maps regular orbits
occur as two-parameter families, leading to two-dimensional
regions in frequency space~\cite{OnkLanKetBae2016}.

Of particular importance for the dynamics are so-called resonances
for which the frequencies fulfill
\begin{equation}
  m_1\nu_1 + m_2\nu_2 = n,
\end{equation}
where $m_1,m_2,n \in \mathbb{Z}$ without a common divisor and at least one of
$m_1$ or $m_2$ is not zero.
The order of a resonance is given by $|m_1|+|m_2|$. These resonance lines,
denoted by $m_1:m_2:n$, form a dense resonance web in frequency space.
In Fig.~\ref{fig:freq_reg_tori}, the curve of regular \twoD{} tori in
frequency is interrupted by several gaps. These gaps are related
to resonance lines for which the corresponding
original regular \twoD{} tori are destroyed.

The number of independent resonance conditions for a given frequency pair
$(\nu_1, \nu_2)$ constitute the rank of the resonance
and correspond to different types of dynamics:
\begin{itemize}
\item
If the frequency pair fulfills no resonance conditions, it is
of rank-0.  The motion on the corresponding \twoD{} torus is
quasi-periodic filling it densely.
This corresponds to the dynamics on the regular cylinders
illustrated in Fig.~\ref{fig:ABC3D}.
\item
If only one resonance condition is satisfied, the
resonance is either (a) uncoupled, i.e., $m_1 :0:n$ or $0:m_2 :n$, or
(b) coupled, i.e., $m_1 :m_2 :n$ with both $m_1$ and $m_2$ nonzero.
Such rank-1 resonances correspond to
quasi-periodic motion on a \oneD{} invariant set which consists
either of one component in the case of coupled resonances or
of $m_1$ (or $m_2$) dynamically connected components in the case
of uncoupled resonances.
We would like to point out that this is the most important type
of resonance for stickiness considered in the next section.

In Fig.~\ref{fig:freq_reg_tori} orbits
belonging to the $2:0:1$ resonance, the $0:4:1$ resonance,
and the $1:5:2$ resonance, respectively,
are shown.
Note that the $2:0:1$ resonance provides an example of
regular orbits which do not fulfill
the slice condition \eqref{eq:slice-condition},
and are therefore not visible in Fig.~\ref{fig:2D-slice}.

\item
For rank-2 resonances, two independent resonance
conditions are fulfilled simultaneously.
While there are a few examples of such double resonances
for the ABC map, they appear to be of no relevance
for the stickiness of orbits considered in the next section.
\end{itemize}

Due to the resonances, which form a dense set of lines in frequency
space, the one-parameter family of \twoD{} tori
has infinitely many holes. Typically, resonances of
low order lead to large gaps, such as the $2:0:1$ resonance.
To the right and below the frequencies
$(\nu^\text{bc}_1, \nu^\text{bc}_2)$ of the boundary cylinder,
all regular tori have to be resonant, as any non-resonant
\twoD{} torus would form a larger boundary cylinder.
Numerically this is not easy to confirm, in particular for large orders.
Moreover, as mentioned above, there could also be orbits for which the
frequency criterion~\eqref{eq:freq-delta} is fulfilled,
but which escape to the chaotic region for very long times.

\section{Stickiness and Power-law Trapping}
\label{sec:stickiness}

To characterize transport and stickiness in a mixed phase space, a
powerful approach is based on the Poincar\'e recurrence theorem. It states that
for a measure-preserving map with invariant probability measure $\mu$, almost
all orbits started in a region $\Lambda$ in phase space will
return to that region at a later time~\cite{Poi1885}.
It is numerically convenient to consider the cumulative
Poincar\'e recurrence statistics defined as
\begin{equation}
  P(t) = \frac{N(t)}{N(0)},
\end{equation}
where $N(0)$ is the number of orbits initially started in $\Lambda$ and
$N(t)$ is the number of orbits which have not yet returned to $\Lambda$
until time $t>0$.
By definition, $P(0) = 1$ and $P(t)$ decreases monotonically with time.

\begin{figure}[h]
  \centering
  \includegraphics{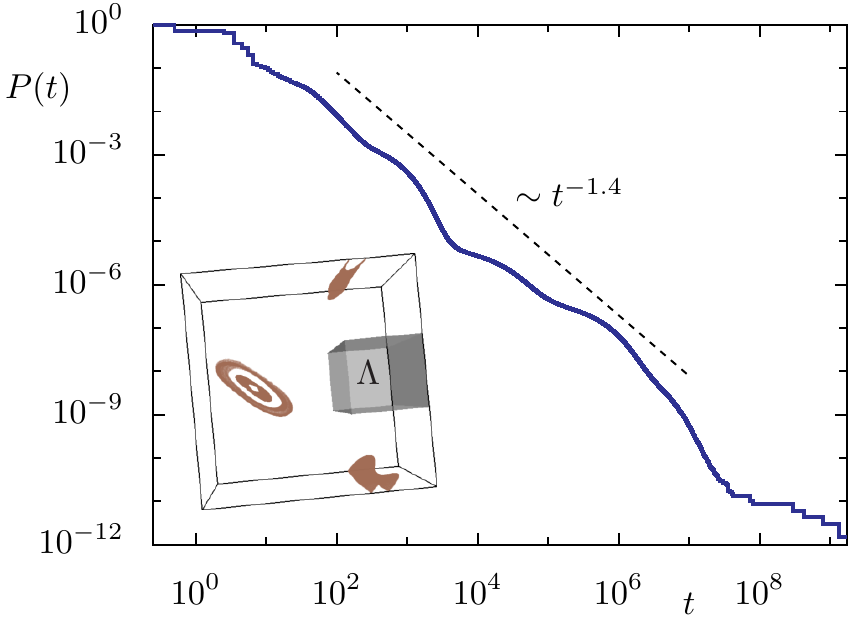}
  \caption{Poincar\'e recurrence statistics $P(t)$ for the ABC
    map~\eqref{equ:ABC} for $(A,B,C) = (2.0,1.5,0.06)$.  The dashed line
    indicates a power-law decay $\sim t^{-\gamma}$ with $\gamma = 1.4$.
    The inset shows the phase space as in Fig.~\ref{fig:ABC3D}(b)
    with the initial region $\Lambda$,
    see Eq.~\eqref{eq:initial_region},
    as grey box.
  }
  \label{fig:power_law_decay}
\end{figure}

\subsection{Poincar\'e recurrence statistics $P(t)$}
\label{subsec:P_t}

The nature of the decay of $P(t)$ is determined by the dynamical properties
of a given system. For a fully chaotic system, the decay is usually
exponential~\cite{BauBer1990,  ZasTip1991, HirSauVai1999, AltSilCal2004}.
However, generic Hamiltonian systems with a mixed phase
space display anomalous transport and typically exhibit much
slower power-law decay~\cite{ChaLeb1980, ChiShe1983,
  Kar1983, ChiShe1984, KayMeiPer1984a, HanCarMei1985, MeiOtt1985,
  ZasEdeNiy1997, BenKasWhiZas1997, ChiShe1999,
  ZasEde2000, WeiHufKet2003, CriKet2008,
  CedAga2013, AluFisMei2014, LanRicOnkBaeKet2014, SilBeiMan2015,
  AluFisMei2017, FirLanKetBae2018}.

\begin{figure}[b]
  \includegraphics{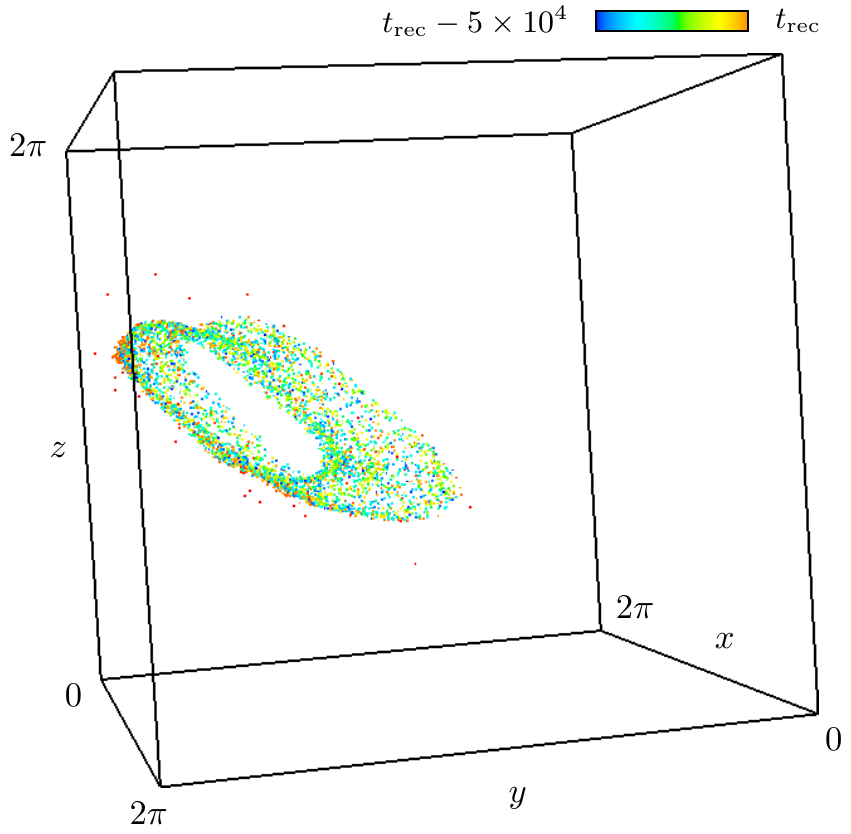}
  \caption{Long-trapped orbit with a recurrence time
    $t_{\text{rec}} = 3\times10^8$ in
    the \threeD{} phase space of the ABC map~\eqref{equ:ABC}.
    This orbit is trapped close to the surface of the boundary cylinder.
    Only the last $5\times 10^4$ iterates are shown with time encoded
    in color.
  }
  \label{fig:long_trapped_3D}
\end{figure}

\begin{figure}[b]
  \includegraphics{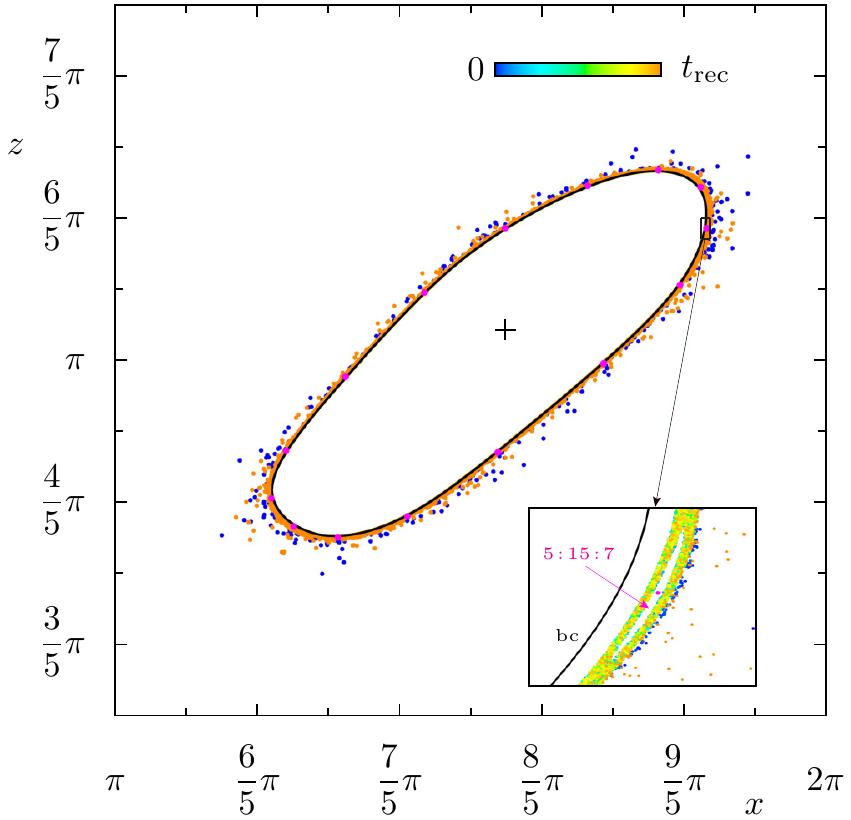}\\
  \caption{Same long-trapped orbit as in Fig.~\ref{fig:long_trapped_3D}
    displayed in the \twoD{} slice $y=\pi$,
    i.e.\ only those points $(x_n, z_n)$ are shown
    for which $|y_n - \pi|< 10^{-3}$ with time encoded in color.
    The black curve shows the boundary cylinder in the \twoD{} slice.
    The location of the central \oneD{}
    torus in indicated by $+$.
    The $5:15:7$ resonance
    appears as 15 resonance islands, enclosed by the trapped orbit,
    as seen by the magnification in the inset.}
  \label{fig:long_trapped_2d_slice}
\end{figure}

\begin{figure}[t]
  \includegraphics{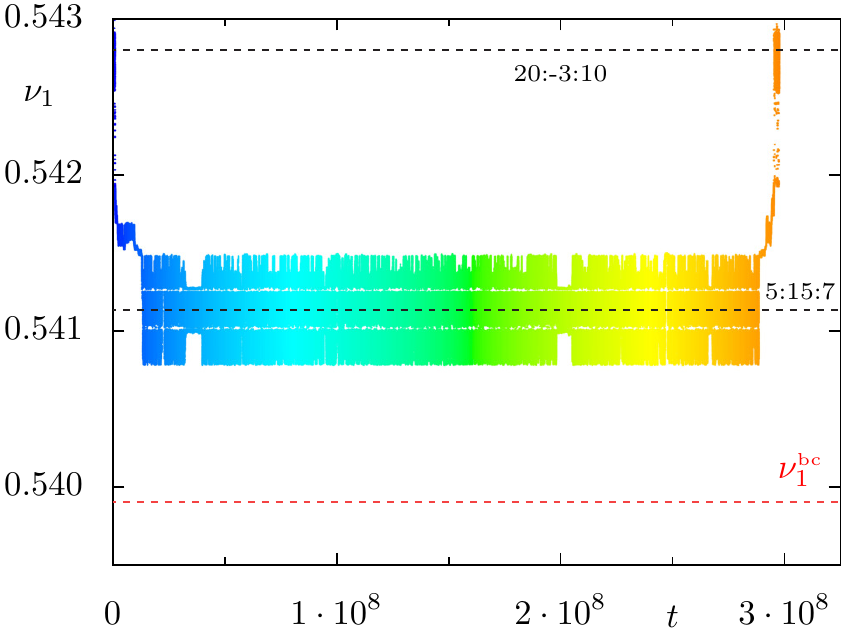}

  \caption{Long-trapped orbit with a recurrence time
    $t_{\text{rec}} = 3\times10^8$ in frequency-time representation. This orbit
    stays close to the coupled resonance $5:15:7$ for most of the time.
    The dashed red line indicates $\nu^\text{bc}_1$
    of the boundary cylinder. Additionally, the $20:-3:10$ resonance is shown,
    which is one of the most dominant ones,
    see Sec.~\ref{subsec:approach_bc}.}
  \label{fig:long_trapped_freq_rep}
\end{figure}

For the ABC map, we choose the following initial region
inside the chaotic sea
\begin{align}\label{eq:initial_region}
  \Lambda = \bigg\{(x,y,z): \Big[0, \frac{2\pi}{3}\Big],[0, 2\pi],
                   [\pi-1, \pi+1] \bigg\}.
\end{align}
For the determination of the Poincar{\'e} recurrence statistics,
we choose $N(0) = 10^{12}$
initial conditions randomly from a uniform distribution in $\Lambda$.
Figure~\ref{fig:power_law_decay} shows the Poincar\'e recurrence
statistics $P(t)$ for the system at $C = 0.06$.
An overall power-law decay is found, with several
significant oscillations superimposed. We will explain their origin
in Sec.~\ref{subsec:P_t_splitting}.
Note that the same type of behavior is also found
for other values of $C\in[0, 1]$ (not shown).

The origin of the slow decay of $P(t)$ are chaotic
orbits which spend long times at either of the two regular regions.
As both regular regions are related by symmetry,
the contributions to $P(t)$ should statistically be the same.
To confirm this, we classify all long-trapped orbits
to belong to either of the regions using
the average value of their $x$ coordinate.
Both types of orbits give identical contributions to $P(t)$ for $t>10^4$,
while initial differences are due the non-symmetric
location of $\Lambda$.
Therefore, in the following analysis of long-trapped orbits,
we restrict to those orbits which belong to the central region.

\subsection{Long-trapped orbits}
\label{subsec:long_trapped_example}

\begin{figure}[b]
   \includegraphics{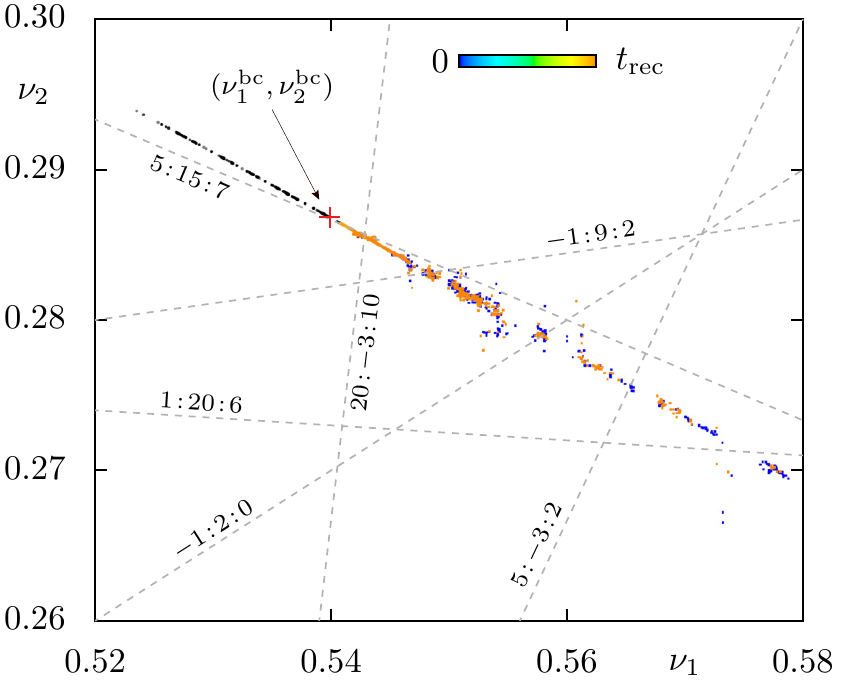}

   \caption{Long-trapped orbit with
            $t_{\text{rec}} = 3\times 10^8$ displayed in frequency
            space with color encoding time.
            A set of $5\times10^6$ pairs of frequencies representing
            regular \twoD{} tori is shown in  black.
            The orbit approaches the boundary cylinder
            $(\nu^{\text{bc}}_1, \nu^{\text{bc}}_2)$
            which is indicated by `+' in red,
            and is trapped most of the time around the $5:15:7$ resonance.}
   \label{fig:freq_space_longest_orbit}
\end{figure}

Long-trapped orbits are found to stay close to the surface of the
boundary cylinder which constitutes an impenetrable barrier.
As representative example we consider in the following
the longest trapped orbit found in our numerical studies
with $t_{\text{rec}} = 3\times 10^8$.
Figure~\ref{fig:long_trapped_3D} shows the last 50 000 iterates of this
orbit in the \threeD{} phase space with time encoded in color. It fills a
cylindrical surface outside of the boundary cylinder before moving away to
the chaotic sea, as seen by the red scattered dots.

The approach to the boundary cylinder is more
clearly seen in the \twoD{} slice by
plotting all points of the long-trapped orbit fulfilling the
slice condition \eqref{eq:slice-condition}, see
Fig.~\ref{fig:long_trapped_2d_slice}.
The magnification in the inset reveals that a small
region is left out, which is due to the
$5:15:7$ resonance.
In the \threeD{} phase space the corresponding elliptic \oneD{} torus
completes 15 rotations on the $x-z$ plane around the
central \oneD{} torus while it goes around 5 times in the $y$-direction.
This therefore leads to 15 small resonance islands in the \twoD{} slice.

To determine the boundary cylinder, shown as black curve in
the \twoD{} slice in Fig.~\ref{fig:2D-slice} and
in Fig.~\ref{fig:long_trapped_2d_slice},
we consider several initial conditions along a line parallel to
the $z$-axis starting from the central \oneD{} torus,
see  Fig.~\ref{fig:long_trapped_2d_slice}.
For each initial condition the frequencies $(\nu_1, \nu_2)$
for the first 4096 points and $(\tilde{\nu}_1, \tilde{\nu}_2)$
for 4096 points after $10^9$ iterates are computed.
The last initial condition for which the
frequency criterion~\eqref{eq:freq-delta}
is fulfilled provides a good approximation to the boundary
cylinder, for which we get
$(\nu^{\text{bc}}_1, \nu^{\text{bc}}_2) \approx (0.5399  , 0.2868)$.
Note that this numerical approach of course cannot guarantee
that this really is a \twoD{} torus as for very large
number of iterations it could eventually escape
towards the chaotic region.

For a long-trapped orbit,
such as shown in Fig.~\ref{fig:long_trapped_3D}
and Fig.~\ref{fig:long_trapped_2d_slice},
the motion is close to the boundary cylinder.
Therefore one can associate finite-time frequencies
by computing a sequence of frequency pairs $(\nu_1^k, \nu_2^k)$, each
determined
from 4096 consecutive points of the orbit.
As the regular tori form a \oneD{} line in frequency space,
see Fig.~\ref{fig:freq_reg_tori},
the approach to the boundary cylinder can
be quantified by either $\nu_1$ or $\nu_2$. Here we use $\nu_1$ which
corresponds to the motion in $y$-direction.

Figure~\ref{fig:long_trapped_freq_rep}
shows for the long-trapped orbit the frequencies $\nu_1^k$
as function of time.
In addition the frequency $\nu^{\text{bc}}_1$ of the boundary cylinder
and the most relevant resonance line $5:15:7$
is shown.
Crossing the $20:-3:10$ resonance,
the orbit rapidly moves towards the frequency of
the boundary cylinder and stays around the resonance line
for most of the time. The same orbit is shown in frequency space
in Fig.~\ref{fig:freq_space_longest_orbit}
with points colored according to time.
While the initial (blue) and final (orange) part extend
over some more scattered range in both $\nu_1$ and $\nu_2$,
most of the time is spend close to $(\nu^{\text{bc}}_1, \nu^{\text{bc}}_2)$.
In addition several gaps are visible which
are related to resonance lines, as we will discuss in more
detail in the next section.

Note that to associate a frequency $\nu_1$ with a resonance line
we use the fact that the regular \twoD{} tori
form an essentially one-dimensional curve
in frequency space, as seen in Fig.~\ref{fig:freq_reg_tori}.
Thus this line intersects
a given resonance line $m_1:m_2:n$ only in one point
which provides a unique $\nu_1$, as displayed
in Fig.~\ref{fig:long_trapped_freq_rep}.
Numerically we fit the data for the regular \twoD{} tori
by a quadratic polynomial which is then used to compute
the intersection points.

\begin{figure*}
  \begin{minipage}{\textwidth}
    \includegraphics[width=1.0\textwidth]{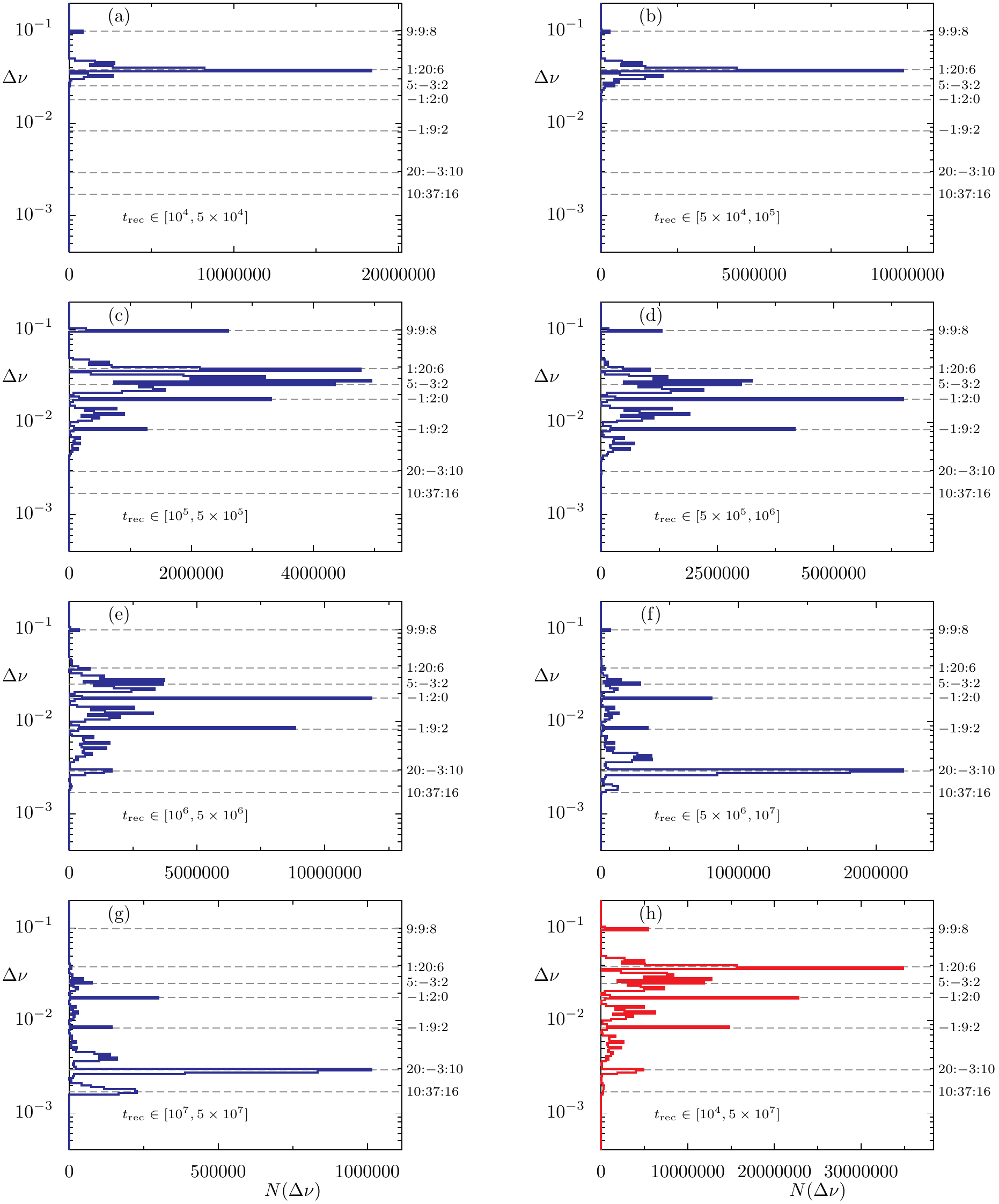}

    \caption{Histogram $N(\Delta\nu)$ of $\Delta\nu = \nu_1 - \nu^{\text{bc}}_1$
      for all long-trapped orbits with recurrence time $t_{\text{rec}} > 10^4$,
      divided in seven time windows:
      (a)
      $t_{\text{rec}} \in [10^4, 5\times10^4]$, (b)
      $t_{\text{rec}} \in [5\times10^4, 10^5]$, (c)
      $t_{\text{rec}} \in [10^5, 5\times10^5]$, (d)
      $t_{\text{rec}} \in [5\times10^5, 10^6]$, (e)
      $t_{\text{rec}} \in [10^6, 5\times10^6]$, (f)
      $t_{\text{rec}} \in [5\times10^6, 10^7]$, (g)
      $t_{\text{rec}} \in [10^7, 5\times10^7]$, and (h)
      $t_{\text{rec}} \in [10^4, 5\times10^7]$ (all).
      This sequence of plots illustrates the approach to
      the boundary cylinder.
      Some relevant coupled resonance lines are also
      displayed.
      These resonances become prominent in different time
      windows and play a significant role in the trapping.
    }
    \label{fig:freq_histogram}
  \end{minipage}
\end{figure*}

\subsection{Approach to the boundary cylinder}
\label{subsec:approach_bc}

Based on the set of long-trapped orbits determined
using the Poincar\'e recurrence statistics we now
analyze their statistical properties.
In particular we identify the most relevant resonance
lines and then use these to split the contributions to $P(t)$
to explain the overall power-law behavior and the superimposed
oscillations.

\begin{figure}[t]
    \includegraphics{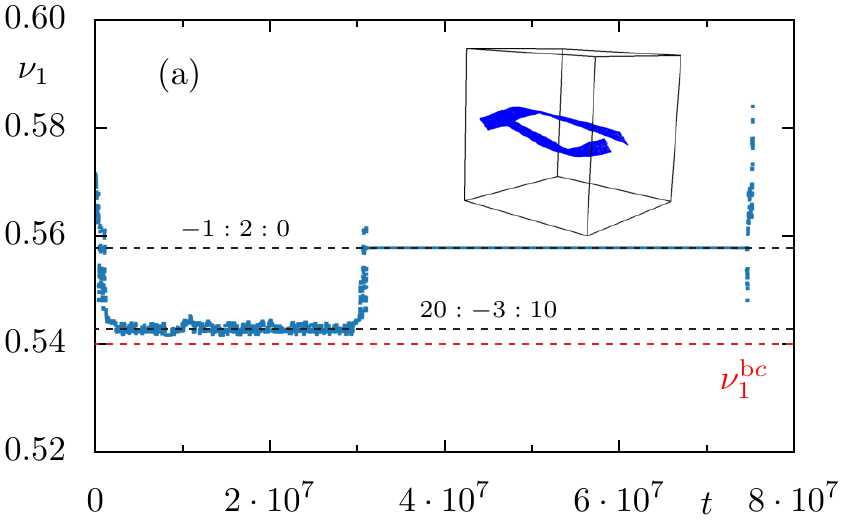}

    \vspace*{1ex}
    \includegraphics{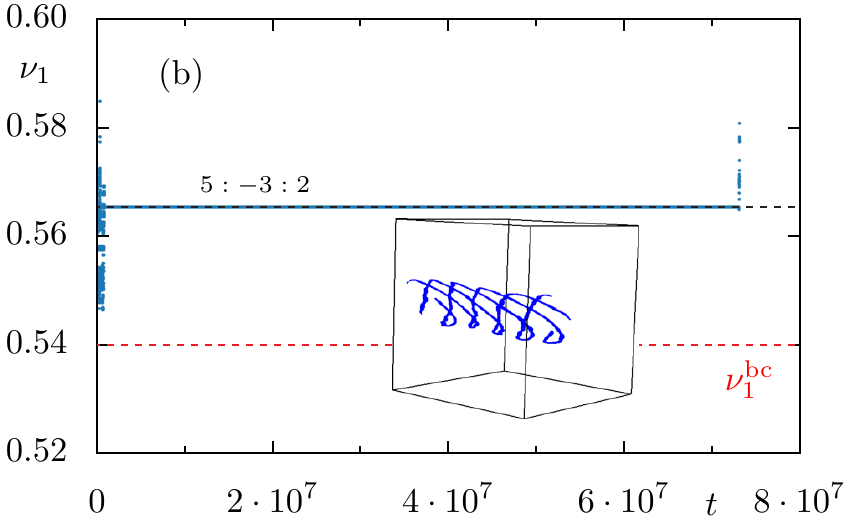}

    \caption{Two examples of long-trapped orbits in frequency-time
        representation. (a) Orbit with $t_{\text{rec}} = 2.2 \times 10^7$.
        The orbit initially approaches the boundary cylinder followed by
        sticking successively to the coupled resonances $20:-3:10$ and
        $-1:2:0$. The inset shows a part of the orbit corresponding to the
        $-1:2:0$ resonance in the \threeD{} phase space. (b)
        Orbit with
        $t_{\text{rec}} = 7.3 \times 10^7$ which sticks for a very
        long time to the coupled resonance $5:-3:2$. The inset illustrates
        the corresponding dynamics in phase space.}
    \label{fig:long_trapped_freq_signals}
\end{figure}

\subsubsection{Identification of resonance lines}

To identify the most relevant resonance lines, a
set of all resonance lines up to order 25 is generated in the
rectangle $\nu_1 \in [0.51, 0.59]$, $\nu_2 \in [0.26,0.30]$
in frequency space.
For one given long-trapped orbit the sequence
of frequency pairs $(\nu_1^k, \nu_2^k)$ is computed
and for each frequency pair the distance
to a resonance line $m_1:m_2:n$ is determined by
\begin{equation}
  \sigma^k = |m_1\nu_1^k + m_2\nu_2^k -n|.
\end{equation}
The resonance with the smallest distance less than a threshold
(set here as $10^{-9}$) is associated with the $k$-th orbit segment.
The resulting six most dominant resonance lines with which
most of the frequency pairs of all orbits are associated
are
-- $9:9:8, 1:20:6, 5:-3:2, -1:2:0, -1:9:2,$ and
$20:-3:-10$. Interestingly, all of them are coupled resonances.
These lines are shown in frequency space
in Fig.~\ref{fig:freq_space_longest_orbit}.

\newcommand{\VSP}{%
{\rule{0pt}{3ex}}        
{\rule[-1.2ex]{0pt}{0pt}}  
}

\begin{table}[h]
  \begin{center}
    \begin{tabular}{ | m{7em} | m{8em}| m{8em}|}
      \hline
      Time window & Number of orbits \VSP\\
      \hline
      $[10^4, 5\times10^4]$ & 1153386  \VSP\\
      \hline
      $[5\times10^4, 10^5]$ & 200898  \VSP\\
      \hline
      $[10^5, 5\times10^5]$ & 103392  \VSP\\
      \hline
      $[5\times10^5, 10^6]$ & 29922   \VSP\\
      \hline
      $[10^6, 5\times10^6]$ & 21298  \VSP\\
      \hline
      $[5\times10^6, 10^7]$ & 749   \VSP\\
      \hline
      $[10^7, 5\times10^7]$ & 167  \VSP\\
      \hline
    \end{tabular}

    \caption {Long-trapped
      orbits categorized in seven time windows (left column) according
      to their return times and the corresponding number (right column).
      } \label{table:Table_1}
  \end{center}
\end{table}

When chaotic orbits approach the boundary cylinder,
these most significant resonances are successively
accessed.  The approach is characterized by the distance
to the boundary cylinder,
\begin{equation}
  \Delta \nu \equiv \Delta \nu^k = |\nu_1^k -\nu^{\text{bc}}_1|.
\end{equation}
All computed long-trapped orbits with  $t_\text{rec} \ge 10^4$
are categorized in seven time windows according to their return times,
see Tab.~\ref{table:Table_1}.
For each of these time windows, we
plot a histogram of  $\Delta\nu$, see Fig.~\ref{fig:freq_histogram}.
One clearly sees that overall small values of the distance $\Delta \nu$
to the boundary cylinder correspond to large recurrence times.
Moreover, the histograms show strong peaks at the
selected dominant resonances lines (shown as dashed gray lines).
Their importance changes with increasing time.
For example, initially for (a) $t_\text{rec}\in[10^4, 5 \times 10^4]$
and (b) $t_\text{rec}\in[5\times10^4, 10^5]$
the $1:20:6$ resonance is most relevant,
while the $-1:2:0$ becomes important for intermediate
time scale, and from $t_\text{rec}=5 \times 10^6$ on
the $20:-3:10$ resonance dominates.
Still, even for orbits dominantly trapped near the
$20:-3:10$ resonance, they may spend significant times
near any of the other resonances with larger $\Delta \nu$,
i.e.\ further away from the boundary cylinder.
Moreover, as seen in Fig.~\ref{fig:freq_histogram}(c)
there can be resonances with larger $\Delta \nu$
playing a significant role for some intermediate times
even if the majority of the orbit is already
trapped with smaller $\Delta \nu$. In addition, a higher order resonance
$10:37:16$ appears to be relevant only for orbits  $t_\text{rec}\in[10^7, 5
\times 10^7]$ in Fig.~\ref{fig:freq_histogram}(g).
In fact, the longest-trapped orbit shown
in Fig.~\ref{fig:long_trapped_freq_rep} is initially trapped around
this resonance for about $10^7$ iterations before
it approaches the $5:15:7$ resonances.

\begin{figure}[t]
    \includegraphics{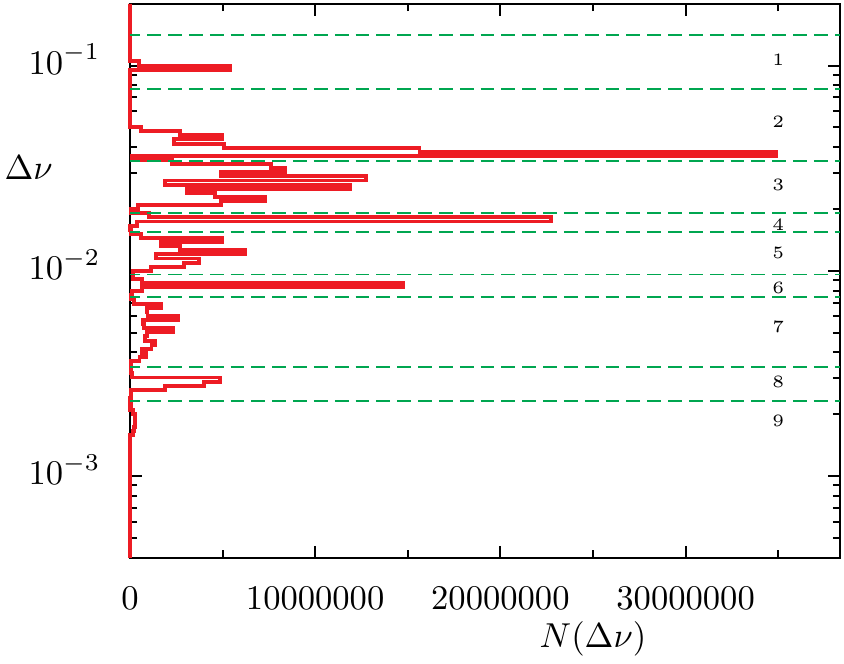}

    \caption{Histogram $N(\Delta\nu)$ of $\Delta \nu = \nu_1 - \nu^{\text{bc}}_1$
      of all orbits with $t_{\text{rec}} \in [10^4, 5\times10^7]$
      divided in 9 different
      chambers by visual inspection.}
    \label{fig:freq_hist_with_chamber}
\end{figure}

\subsubsection{Examples of coupled resonance}

Figure~\ref{fig:long_trapped_freq_signals} shows
the frequency $\nu_1$ as function of time for two examples
illustrating the relevance and geometry of coupled resonances.
In Fig.~\ref{fig:long_trapped_freq_signals}(a)
an orbit with $t_{\text{rec}} = 2.2 \times 10^7$ is shown,
which initially approaches $\nu^{\text{bc}}_1$
and fluctuates around the frequency of the $20:-3:10$
resonance.
Finally, the orbits sticks on the coupled $-1:2:0$ resonance for an even longer time. The corresponding
dynamics in the \threeD{} phase space is illustrated in the inset
and consists of two separate, but dynamically connected parts.
In Fig.~\ref{fig:long_trapped_freq_signals}(b)
an orbit with return time $t_{\text{rec}}= 7.3\times 10^7$
is shown. Here $5:-3:2$ is the dominant resonance.
The corresponding dynamics in the \threeD{} phase space
shown in the inset spirals five times around in the $y$-direction
and three times around the central \oneD{} torus.
For both orbits $\nu_1$ shows on this scale essentially no fluctuations
around the frequency of the corresponding resonance.
A closer inspection using the representation in the \twoD{} slice
shows no indication that higher order resonances deeper
in the class hierarchy around the corresponding resonance islands
are accessed. Still we expect that this is possible
for even longer trapped orbits.

\subsection{Splitting of $P(t)$}
\label{subsec:P_t_splitting}

\begin{figure}[h]
  \includegraphics{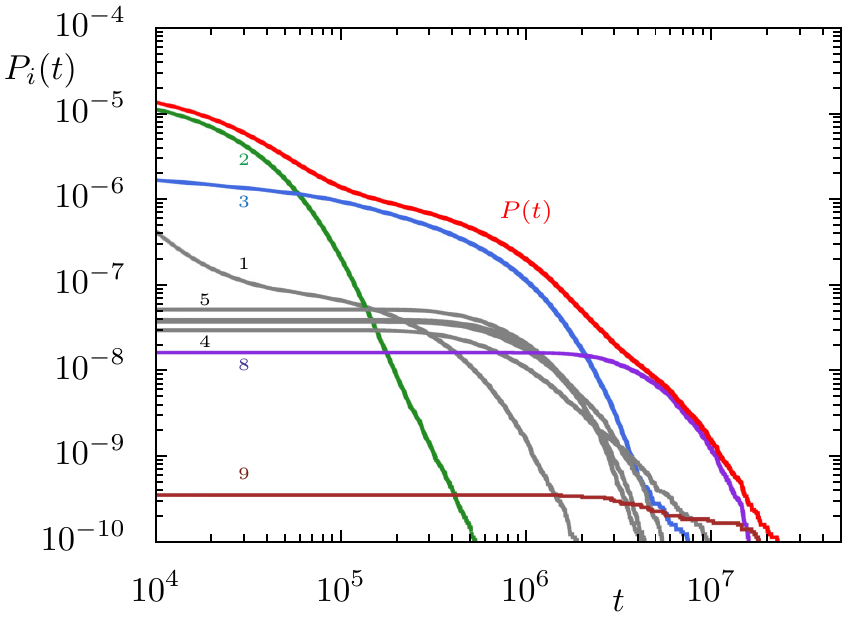}

  \caption{Poincar\'e recurrence statistics $P(t)$ split in
    contributions $P_i(t)$ of the chambers identified
    in Fig.~\ref{fig:freq_hist_with_chamber}.
    For comparison, the full $P(t)$ is shown in red.
    }
    \label{fig:P_t_split_all_chambers}
\end{figure}

The plot in Fig.~\ref{fig:freq_histogram}(h) shows the complete distribution of
$\Delta\nu$ for all long-trapped orbits for
$t_{\text{rec}} \ge 10^4$.
The minima of this distribution indicate frequencies
which should correspond to most restrictive partial barriers.
Based on visual inspection, we classify this distribution
in nine chambers as displayed in
Fig.~\ref{fig:freq_hist_with_chamber}.
The aim is to quantify the contributions to the
Poincar\'e recurrence statistics $P(t)$
from each of these chambers.
For this we associate, based on $\Delta\nu$, with each long-trapped orbit
the most relevant chamber, i.e.\
the one for which it spends the longest time.
By this the total $P(t)$ is split into
contributions $P_i(t)$, determined from those orbits associated
with the $i$-th chamber.
Figure~\ref{fig:P_t_split_all_chambers} shows these contributions
together with the full $P(t)$.
Several of the $P_i(t)$ show an initially exponential decay.
In contrast, for $P_2(t)$ an overall power-law is found for $t\gtrsim 10^5$.
This suggests that there is some generalized
hierarchy (e.g.\ involving the equivalent
of the class hierarchy in \twoD{} maps \cite{Mei1986}),
which is accessed by orbits in this chamber.
Moreover, chamber 2 also contributes significantly in the initial decay
up to $t\simeq 7 \times 10^4$
of the full $P(t)$ shown in red.

\begin{figure}[h!]
  \includegraphics{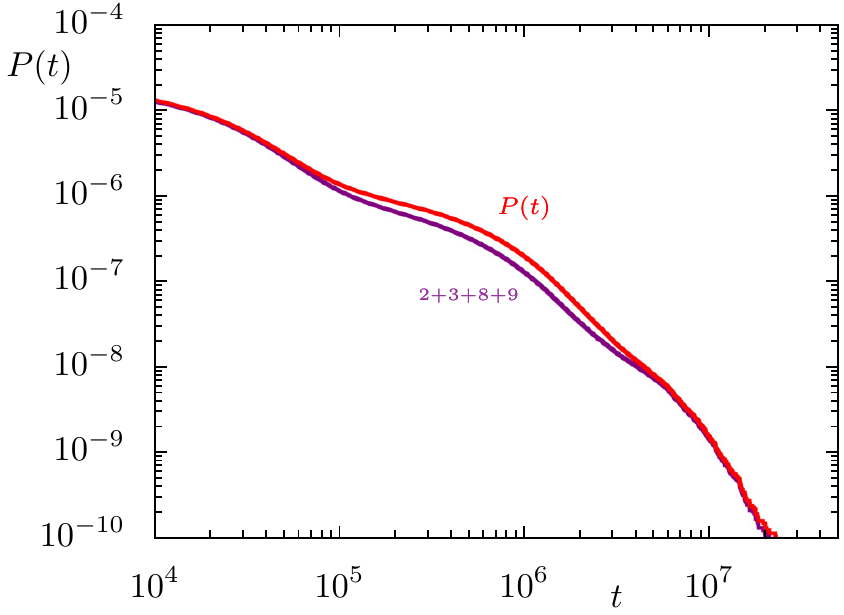}

  \caption{The full Poincar\'e recurrence statistics $P(t)$
           is well approximated by Eq.~\eqref{eq:p_t_sum}, i.e.\
         by the sum of only four contributions $P_i(t)$
         with $i= 2, 3, 8, 9$,
         as identified in Fig.~\ref{fig:freq_hist_with_chamber}.}
  \label{fig:P_t_split_important_chambers}
\end{figure}

From $t\simeq 7 \times 10^4$ on the most relevant contribution
comes from chamber 3 for which $P_3(t)$ shows good agreement
with an exponential decay,
as manifested in a semi-logarithmic plot
giving a straight line over several orders of magnitude
(not shown).
Starting from approximately $t=3\times 10^6$
one sees that $P_8(t)$ gives the most important contribution,
again well-described by an exponential.
In particular, the sum of the four most important $P_i(t)$
provides a good approximation of the full $P(t)$,
\begin{equation}
\label{eq:p_t_sum}
  P(t) \approx P_2(t) + P_3(t) + P_8(t) + P_9(t),
\end{equation}
see Fig.~\ref{fig:P_t_split_important_chambers}.
This therefore on the one hand explains
the overall power-law behavior of $P(t)$ as
sum of exponentially decaying contributions, in the spirit
of the explanation of power-law trapping
in \twoD{} systems  \cite{HanCarMei1985}.
On the other hand, a power-law requires
a particular scaling of the individual contributions
and is only expected to hold asymptotically for large $t$.
The observed superimposed oscillations
can be traced back to the individual contributions $P_i(t)$.
In particular $P_3(t)$ is responsible for the prominent hump near $t=10^6$.
In general it has been shown that
uncorrelated fluctuations in the
transition rates between regions separated by partial barriers
lead to slowly decaying oscillations in $P(t)$ \cite{CedAga2013}.

\subsection{Partial barriers}
\label{sec:partial-barriers}

Overall, the mechanism of trapping in the \threeD{} ABC map
appears to be similar to the well-known trapping in \twoD{} maps:
Long-trapped orbits approach a boundary-cylinder
and longer trapping times correspond to a closer approach,
which is similar to the level hierarchy \cite{Kay1983}.
Also trapping within resonances is found, which is
similar to the class hierarchy \cite{Mei1986}. In the \twoD{} case
these hierarchies can be arranged in a binary tree
\cite{KayMeiPer1984b,MeiOtt1986},
which under certain statistical assumptions
leads to a universal asymptotic decay \cite{CriKet2008,AluFisMei2017}.
However, an important difference to the \twoD{} case is that
for the \threeD{} ABC map
coupled resonances are possible and appear to
play the key role for long-trapped orbits.
The numerics suggests that there must be some
kind of partial barriers
which define the borders between the chambers
containing  the identified resonances.
Geometrically such partial barriers should correspond to broken
cylindrical \twoD{} tori. Here \twoD{} tori
which are furthest away from resonances should
persist longest with increasing perturbation from the uncoupled case.
The break-up of invariant tori in volume-preserving
maps has been studied recently in \cite{FoxMei2013,FoxMei2016}
showing that tori seem to stretch near break-up which might lead to analogues of gaps in cantori for \twoD{} maps.
Computing a sequence of \oneD{} tori approximating
such partial barriers and determining the transport
through them for volume-preserving maps such as the ABC map,
is an interesting non-trivial task left for future studies.

\section{Summary and outlook}
\label{sec:summary}

In this paper, we study stickiness in volume-preserving maps using
the example of the ABC map. The \threeD{} phase space of the map is mixed with
two disjoint regular regions embedded in a chaotic sea. The map
reduces to a (2+1)\someD{} system when $C=0$. In this case the motion in the
$y$-direction decouples from the \twoD{} area-preserving $x-z$ subsystem,
which has two elliptic fixed points.
For the uncoupled case regular regions of the \threeD{} map are in the form of
concentric cylinders organized around \oneD{} elliptic tori corresponding
to the fixed points of the \twoD{} subsystem.
With increasing $C>0$ these cylinders are deformed.
As long as these deformed cylinders acquire no holes,
they form absolute barriers to the motion.
If a cylinder is destroyed, this is expected to lead to a
partial barrier to transport for chaotic orbits.
For a finite $C$, destruction of \twoD{} tori around \oneD{} tori
may occur inside
the regular regions in addition to the interface of the
chaotic and regular region outside.
Thus chaotic layers appear inside the regular
regions which are not yet connected to the chaotic sea,
as is nicely seen in the sequence of plots of the
finite-time Lyapunov exponents in Fig.~\ref{fig:Lyap_exp_plot}.
Representing the \twoD{} regular tori in frequency
space shows that they lead to an essentially one-dimensional curve
which is attached to the frequencies of the
central elliptic \oneD{} torus. Resonances lead
to gaps within the family of \twoD{} tori.
An outermost regular torus is identified as boundary cylinder
which provides the separation to the chaotic sea.

Using the cumulative Poincar\'e recurrence statistics $P(t)$
allows to investigate the stickiness of chaotic orbits.
For the ABC map we find an overall power-law decay
with significant oscillations superimposed.
Long-trapped orbits approach the boundary cylinder
and their segment-wise frequency-time representation
reveals that they spend long times near six
most relevant resonances.
Using the distribution of all frequencies of all long-trapped
orbits shows that they can be classified into nine chambers.
Each long-trapped orbit is classified according
to the chamber in which it spends the largest amount of time.
From this we obtain a splitting of the
cumulative Poincar\'e recurrence statistics $P(t)$
into different contributions $P_i(t)$.
These show good agreement with exponential decays and
the sum of only four of them gives an excellent approximation
to the full $P(t)$.
This explains both the overall approximate power-law decay,
as a superposition of individual exponential decays,
and also the superimposed oscillations due
to the individual $P_i(t)$.
This also suggests the existence of some kind of partial barriers
separating the different chambers. Most likely,
these are broken \twoD{} tori, similar to cantori
in \twoD{} maps.

An important task for the future is
to explicitly determine the partial barriers, e.g.\
by sequences of approximating families of \oneD{} tori,
and to identify the turnstile and to
determine the corresponding flux.
This should eventually allow for setting up a quantitative
statistical prediction of the observed power-law trapping.
Moreover, a fundamental question concerns
whether there is a universal value of the exponent
$\gamma$ of the power-law decay, similar to that in
area-preserving maps.
In addition, power-law trapping indicates anomalous diffusion,
so that a detailed study of the diffusive properties
in the ABC map would be very interesting.

~

\begin{acknowledgments}
  We are grateful for discussions
  with Markus Firmbach, Franziska H\"ubner, Roland Ketzmerick,
  and Jim Meiss.
  Furthermore, we acknowledge support
  by the Deutsche Forschungsgemeinschaft
  under grant KE~537/6--1.

  All \threeDD{} visualizations were created using
  \textsc{Mayavi}~\cite{RamVar2011}.
\end{acknowledgments}

\bibliographystyle{cpg_unsrt_title_for_phys_rev}

\bibliography{abbrevs,4D_maps,accelerator_modes,arnold_diffusion,baecker,bec,bifurk,billiards,bleh_lit,books_on_chaos,branched_flows,celestial,chemistry,classm,cold_atoms,complexified_systems,continued_fractions,coupled_billiards,CP_theses,desy_pp,diffr,diffusion,dissipative_maps,div_lit,driven_coulomb,eigenval,emerging,entanglement,erg,ev_statistics,exceptional_points,experiments,extreme_events,floquet,fractals,fractal_weyl,frequency_analysis,fun,geholt,general,Hamiltonian_systems,higher_dim,higher_dim_bif,higher_dim_billiards,higher_dim_manifolds,hreg_lie_canon_trans,inside_outside,interessant_nicht_geholt,libraries,lyapunov_exponents,manifolds,many_body,maps,matrix_models,mesoscopic,misc_jb,mjk_unsrt,naff,near_integrable,negcurv,neu,nhim,NMsRandomFindings,nn_lit,noninvertible_maps,n_particle,numerical_algorithms,Numerics,open_systems,part_barr,pot,prim_lit,pseudoint,pt_symmetry,quantized_maps,quantum_transport,random_waves,recurrence_time_distribution,resonances,resonance_trapping,return_times,rob_lit,scars,scientific_computing,sd_beta,shadowing,stadion,statistical,steven_creagh,stochastic_maps,symbd,synchronization,time_dev,time_evolution,time_series_analysis,torus_bifurcations,transport_in_Hamiltonian_systems,tunneling,velocities,verlorene,volume_preserving_maps,wave,chimera}

\begin{thebibliography}{10}
\newcommand{\enquote}[1]{``#1''}
\providecommand{\url}[1]{\texttt{#1}}
\providecommand{\urlprefix}{URL }
\providecommand{\eprint}[2][]{\url{#2}}

\bibitem{AreBlaBudCarCarCleElFeuGolGouvanKraLeMacMelMetMezMouPirSpeStuThiTuv2017}
H.~Aref, J.~R. Blake, M.~Budi{\v s}i\'c, S.~S.~S. Cardoso, J.~H.~E. Cartwright,
  H.~J.~H. Clercx, K.~El~Omari, U.~Feudel, R.~Golestanian, E.~Gouillart, G.~F.
  {van Heijst}, T.~S. Krasnopolskaya, Y.~Le~Guer, R.~S. MacKay, V.~V. Meleshko,
  G.~Metcalfe, I.~Mezi\'c, A.~P.~S. {de Moura}, O.~Piro, M.~F.~M. Speetjens,
  R.~Sturman, J.-L. Thiffeault, and I.~Tuval, \emph{Frontiers of chaotic
  advection}, Rev.~Mod.~Phys. \textbf{89}, 025007 (2017).

\bibitem{She2010}
D.~L. Shepelyansky, \emph{Poincar\'e recurrences in {Hamiltonian} systems with
  a few degrees of freedom}, Phys.~Rev.~E \textbf{82}, 055202(R) (2010).

\bibitem{SetKes2012}
A.~Sethi and S.~Keshavamurthy, \emph{Driven coupled {Morse} oscillators:
  visualizing the phase space and characterizing the transport}, Mol.~Phys.
  \textbf{110}, 717 (2012).

\bibitem{Pap2014}
Y.~{Papaphilippou}, \emph{Detecting chaos in particle accelerators through the
  frequency map analysis method}, Chaos \textbf{24}, 024412 (2014).

\bibitem{OyaSzeBatSouCalViaSan2016}
R.~S. Oyarzabal, J.~D. Szezech, A.~M. Batista, S.~L.~T. {de Souza}, I.~L.
  Caldas, R.~L. Viana, and M.~A.~F. Sanju\'an, \emph{Transient chaotic
  transport in dissipative drift motion}, Phys.~Lett.~A \textbf{380}, 1621
  (2016).

\bibitem{Poi1890}
H.~{Poincar\'e}, \emph{Sur le probl\`eme des trois corps et les \'equations de
  la dynamique}, Acta~Math. \textbf{13}, 1 (1890).

\bibitem{BauBer1990}
W.~Bauer and G.~F. Bertsch, \emph{Decay of ordered and chaotic systems},
  Phys.~Rev.~Lett. \textbf{65}, 2213 (1990).

\bibitem{ZasTip1991}
G.~M. Zaslavsky and M.~K. Tippett, \emph{Connection between recurrence-time
  statistics and anomalous transport}, Phys.~Rev.~Lett. \textbf{67}, 3251
  (1991).

\bibitem{HirSauVai1999}
M.~Hirata, B.~Saussol, and S.~Vaienti, \emph{Statistics of return times: A
  general framework and new applications}, Commun.~Math.~Phys. \textbf{206}, 33
  (1999).

\bibitem{AltSilCal2004}
E.~G. Altmann, E.~C. {da Silva}, and I.~L. Caldas, \emph{Recurrence time
  statistics for finite size intervals}, Chaos \textbf{14}, 975 (2004).

\bibitem{ChaLeb1980}
S.~R. Channon and J.~L. Lebowitz, \emph{Numerical experiments in stochasticity
  and homoclinic oscillation}, Ann.~N.Y.~Acad.~Sci. \textbf{357}, 108 (1980).

\bibitem{ChiShe1983}
B.~V. Chirikov and D.~L. Shepelyansky, \emph{Statistics of {P}oincar\'e
  recurrences and the structure of the stochastic layer of a nonlinear
  resonance},  (1983), in \emph{Proceedings of the 9th International Conference
  on Nonlinear Oscillations, Kiev, 1981; Qualitative methods of the theory of
  nonlinear oscillators}, vol.~2, edited by Yu. A. Mitropolsky, 421--424,
  Naukova Dumka, Kiev, (1984). English translation: Princeton University Report
  No. PPPL-TRANS-133 (1983).

\bibitem{Kar1983}
C.~F.~F. Karney, \emph{Long-time correlations in the stochastic regime},
  Physica~D \textbf{8}, 360 (1983).

\bibitem{ChiShe1984}
B.~V. Chirikov and D.~L. Shepelyansky, \emph{Correlation properties of
  dynamical chaos in {Hamiltonian} systems}, Physica~D \textbf{13}, 395 (1984).

\bibitem{KayMeiPer1984a}
R.~S. MacKay, J.~D. Meiss, and I.~C. Percival, \emph{Stochasticity and
  transport in {Hamiltonian} systems}, Phys.~Rev.~Lett. \textbf{52}, 697
  (1984).

\bibitem{HanCarMei1985}
J.~D. Hanson, J.~R. Cary, and J.~D. Meiss, \emph{Algebraic decay in
  self-similar {Markov} chains}, J.~Stat.~Phys. \textbf{39}, 327 (1985).

\bibitem{MeiOtt1985}
J.~D. Meiss and E.~Ott, \emph{Markov-tree model of intrinsic transport in
  {H}amiltonian systems}, Phys.~Rev.~Lett. \textbf{55}, 2741 (1985).

\bibitem{KanKon1989}
K.~Kaneko and T.~Konishi, \emph{Diffusion in {Hamiltonian} dynamical systems
  with many degrees of freedom}, Phys. Rev. A \textbf{40}, 6130 (1989).

\bibitem{KonKan1990}
T.~Konishi and K.~Kaneko, \emph{Diffusion in {H}amiltonian chaos and its size
  dependence}, J.~Phys.~A \textbf{23}, L715 (1990).

\bibitem{DinBouOtt1990}
M.~Ding, T.~Bountis, and E.~Ott, \emph{Algebraic escape in higher dimensional
  {Hamiltonian} systems}, Phys.~Lett.~A \textbf{151}, 395 (1990).

\bibitem{ChiVec1993}
B.~V. Chirikov and V.~V. Vecheslavov, \emph{Theory of fast {A}rnold diffusion
  in many-frequency systems}, J.~Stat.~Phys. \textbf{71}, 243 (1993).

\bibitem{ChiVec1997}
B.~V. Chirikov and V.~V. Vecheslavov, \emph{Arnold diffusion in large systems},
  J.~Exp.~Theor.~Phys. \textbf{85}, 616 (1997).

\bibitem{ZasEdeNiy1997}
G.~M. Zaslavsky, M.~Edelmann, and B.~A. Niyazov, \emph{Self-similarity,
  renormalization, and phase space nonuniformity of {Hamiltonian} chaotic
  dynamics}, Chaos \textbf{7}, 159 (1997).

\bibitem{BenKasWhiZas1997}
S.~Benkadda, S.~Kassibrakis, R.~B. White, and G.~M. Zaslavsky,
  \emph{Self-similarity and transport in the standard map}, Phys.~Rev.~E
  \textbf{55}, 4909 (1997).

\bibitem{ChiShe1999}
B.~V. Chirikov and D.~L. Shepelyansky, \emph{Asymptotic statistics of
  {Poincar\'e} recurrences in {Hamiltonian} systems with divided phase space},
  Phys.~Rev.~Lett. \textbf{82}, 528 (1999).

\bibitem{ZasEde2000}
G.~M. Zaslavsky and M.~Edelmann, \emph{Hierarchical structures in the phase
  space and fractional kinetics: {I}. {C}lassical systems}, Chaos \textbf{10},
  135 (2000).

\bibitem{WeiHufKet2003}
M.~Weiss, L.~Hufnagel, and R.~Ketzmerick, \emph{Can simple renormalization
  theories describe the trapping of chaotic trajectories in mixed systems?},
  Phys. Rev. E \textbf{67}, 046209 (2003).

\bibitem{AltKan2007}
E.~G. Altmann and H.~Kantz, \emph{Hypothesis of strong chaos and anomalous
  diffusion in coupled symplectic maps}, Europhys.~Lett. \textbf{78}, 10008
  (2007).

\bibitem{ShoLiKomTod2007b}
A.~Shojiguchi, C.-B. Li, T.~Komatsuzaki, and M.~Toda, \emph{Fractional behavior
  in multidimensional {Hamiltonian} systems describing reactions}, Phys.~Rev.~E
  \textbf{76}, 056205 (2007), erratum ibid. {\bf 77}, 019902(E) (2008).

\bibitem{CriKet2008}
G.~Cristadoro and R.~Ketzmerick, \emph{Universality of algebraic decays in
  {H}amiltonian systems}, Phys.~Rev.~Lett. \textbf{100}, 184101 (2008).

\bibitem{CedAga2013}
R.~Ceder and O.~Agam, \emph{Fluctuations in the relaxation dynamics of mixed
  chaotic systems}, Phys.~Rev.~E \textbf{87}, 012918 (2013).

\bibitem{AluFisMei2014}
O.~Alus, S.~Fishman, and J.~D. Meiss, \emph{Statistics of the
  island-around-island hierarchy in {H}amiltonian phase space}, Phys.~Rev.~E
  \textbf{90}, 062923 (2014).

\bibitem{LanBaeKet2016}
S.~Lange, A.~B{\"a}cker, and R.~Ketzmerick, \emph{What is the mechanism of
  power-law distributed {Poincar\'e} recurrences in higher-dimensional
  systems?}, EPL \textbf{116}, 30002 (2016).

\bibitem{AluFisMei2017}
O.~Alus, S.~Fishman, and J.~D. Meiss, \emph{Universal exponent for transport in
  mixed {H}amiltonian dynamics}, Phys.~Rev.~E \textbf{96}, 032204 (2017).

\bibitem{FirLanKetBae2018}
M.~Firmbach, S.~Lange, R.~Ketzmerick, and A.~B\"acker, \emph{Three-dimensional
  billiards: Visualization of regular structures and trapping of chaotic
  trajectories}, Phys.~Rev.~E \textbf{98}, 022214 (2018).

\bibitem{BucDelZakManArnWal1995}
A.~Buchleitner, D.~Delande, J.~Zakrzewski, R.~N. Mantegna, M.~Arndt, and
  H.~Walther, \emph{Multiple time scales in the microwave ionization of
  {R}ydberg atoms}, Phys.~Rev.~Lett. \textbf{75}, 3818 (1995).

\bibitem{BenCasMasShe2000}
G.~Benenti, G.~Casati, G.~Maspero, and D.~L. Shepelyansky, \emph{Quantum
  {P}oincar\'e recurrences for a hydrogen atom in a microwave field},
  Phys.~Rev.~Lett. \textbf{84}, 4088 (2000).

\bibitem{EzrWig2009}
G.~S. Ezra and S.~Wiggins, \emph{Phase-space geometry and reaction dynamics
  near index 2 saddles}, J.~Phys.~A \textbf{42}, 205101 (2009).

\bibitem{MazShe2015}
A.~K. Mazur and D.~Shepelyansky, \emph{Algebraic statistics of {P}oincar\'e
  recurrences in a {DNA} molecule}, Phys.~Rev.~Lett. \textbf{115}, 188104
  (2015).

\bibitem{KayMeiPer1984b}
R.~S. MacKay, J.~D. Meiss, and I.~C. Percival, \emph{{Transport in
  {H}amiltonian systems}}, Physica~D \textbf{13}, 55 (1984).

\bibitem{Mei1986}
J.~D. Meiss, \emph{Class renormalization: Islands around islands}, Phys.~Rev.~A
  \textbf{34}, 2375 (1986).

\bibitem{Mei1992}
J.~D. Meiss, \emph{Symplectic maps, variational principles, and transport},
  Rev.~Mod.~Phys. \textbf{64}, 795 (1992).

\bibitem{AltKan2005}
E.~G. Altmann and H.~Kantz, \emph{Recurrence time analysis, long-term
  correlations, and extreme events}, Phys.~Rev.~E \textbf{71}, 056106 (2005).

\bibitem{MotMouGreKan2005}
A.~E. Motter, A.~P.~S. de~Moura, C.~Grebogi, and H.~Kantz, \emph{Effective
  dynamics in {H}amiltonian systems with mixed phase space}, Phys.~Rev.~E
  \textbf{71}, 036215 (2005).

\bibitem{AltMotKan2006}
E.~G. Altmann, A.~E. Motter, and H.~Kantz, \emph{Stickiness in {H}amiltonian
  systems: From sharply divided to hierarchical phase space}, Phys.~Rev.~E
  \textbf{73}, 026207 (2006).

\bibitem{Mei2015}
J.~D. Meiss, \emph{Thirty years of turnstiles and transport}, Chaos
  \textbf{25}, 097602 (2015).

\bibitem{Arn1964}
V.~I. Arnol'd, \emph{Instability of dynamical systems with several degrees of
  freedom}, Sov.~Math.~Dokl. \textbf{5}, 581 (1964).

\bibitem{Loc1993}
P.~Lochak, \emph{Hamiltonian perturbation theory: periodic orbits, resonances
  and intermittency}, Nonlinearity \textbf{6}, 885 (1993).

\bibitem{Dum2014}
H.~S. Dumas, \emph{The {KAM} Story: {A} Friendly Introduction to the Content,
  History, and Significance of Classical {Kolmogorov}--{Arnold}--{Moser}
  Theory}, World Scientific, Singapore (2014).

\bibitem{MarDavEzr1987}
C.~C. Martens, M.~J. Davis, and G.~S. Ezra, \emph{Local frequency analysis of
  chaotic motion in multidimensional systems: {Energy} transport and
  bottlenecks in planar {OCS}}, Chem.~Phys.~Lett. \textbf{142}, 519 (1987).

\bibitem{Wig1990}
S.~Wiggins, \emph{On the geometry of transport in phase space {I}. {T}ransport
  in $k$-degree-of-freedom {H}amiltonian systems, $2 \leq k < \infty$},
  Physica~D \textbf{44}, 471 (1990).

\bibitem{MacMei1992}
R.~S. MacKay and J.~D. Meiss, \emph{Cantori for symplectic maps near the
  anti-integrable limit}, Nonlinearity \textbf{5}, 149 (1992).

\bibitem{CheSun1990a}
C.-Q. Cheng and Y.-S. Sun, \emph{Existence of invariant tori in
  three-dimensional measure-preserving mappings}, Celest.~Mech.~Dyn.~Astron.
  \textbf{47}, 275 (1990).

\bibitem{CheSun1990b}
C.-Q. Cheng and Y.-S. Sun, \emph{Existence of periodically invariant curves in
  3-dimensional measure-preserving mappings}, Celest.~Mech.~Dyn.~Astron.
  \textbf{47}, 293 (1990).

\bibitem{CarFeiPir1994}
J.~H.~E. Cartwright, M.~Feingold, and O.~Piro, \emph{Passive scalars and
  three-dimensional {Liouvillian} maps}, Physica~D \textbf{76}, 22 (1994).

\bibitem{LomMei1998}
H.~E. Lomel\'{i} and J.~D. Meiss, \emph{Quadratic volume-preserving maps},
  Nonlinearity \textbf{11}, 557 (1998).

\bibitem{Mez2001}
I.~Mezi{\'c}, \emph{Break-up of invariant surfaces in action--angle--angle maps
  and flows}, Physica~D \textbf{154}, 51 (2001).

\bibitem{LomMei2003}
H.~E. Lomel{\'i} and J.~D. Meiss, \emph{Heteroclinic intersections between
  invariant circles of volume-preserving maps}, Nonlinearity \textbf{16}, 1573
  (2003).

\bibitem{LomMei2009b}
H.~E. Lomel{\'i} and J.~D. Meiss, \emph{Generating forms for exact
  volume-preserving maps}, Discrete Contin.~Dyn.~Sys.~Ser.~S \textbf{2}, 361
  (2009).

\bibitem{MirLom2010}
J.~D. Mireles~James and H.~Lomel{\'i}, \emph{Computation of heteroclinic arcs
  with application to the volume preserving {H}\'enon family}, {SIAM}
  J.~Appl.~Dyn.~Syst. \textbf{9}, 919 (2010).

\bibitem{Kor2010}
N.~Korneev, \emph{Perturbation series for calculation of invariant surface
  splitting in volume-preserving maps}, Chaos \textbf{20}, 043102 (2010).

\bibitem{DulMei2012}
H.~R. Dullin and J.~D. Meiss, \emph{Resonances and twist in volume-preserving
  mappings}, {SIAM} J.~Appl.~Dyn.~Syst. \textbf{11}, 319 (2012).

\bibitem{Mei2012}
J.~D. Meiss, \emph{The destruction of tori in volume-preserving maps},
  Commun.~Nonlinear~Sci.~Numer.~Simulat. \textbf{17}, 2108 (2012).

\bibitem{VaiMez2012}
U.~Vaidya and I.~Mezi{\'c}, \emph{Existence of invariant tori in three
  dimensional maps with degeneracy}, Physica~D \textbf{241}, 1136 (2012).

\bibitem{FoxMei2013}
A.~M. Fox and J.~D. Meiss, \emph{Greene's residue criterion for the breakup of
  invariant tori of volume-preserving maps}, Physica~D \textbf{243}, 45 (2013).

\bibitem{Mir2013}
J.~D. Mireles~James, \emph{Quadratic volume-preserving maps: {(Un)stable}
  manifolds, hyperbolic dynamics, and vortex-bubble bifurcations},
  J.~Nonlinear~Sci. \textbf{23}, 585 (2013).

\bibitem{FoxLla2015}
A.~M. Fox and R.~de~la Llave, \emph{Barriers to transport and mixing in
  volume-preserving maps with nonzero flux}, Physica~D \textbf{295--296}, 1
  (2015).

\bibitem{FoxMei2016}
A.~M. Fox and J.~D. Meiss, \emph{Computing the conjugacy of invariant tori for
  volume-preserving maps}, {SIAM} J.~Appl.~Dyn.~Syst. \textbf{15}, 557 (2016).

\bibitem{MaeSmiMit2017}
B.~Maelfeyt, S.~Smith, and K.~Mitchell, \emph{Using invariant manifolds to
  construct symbolic dynamics for three-dimensional volume-preserving maps},
  {SIAM} J.~Appl.~Dyn.~Syst. \textbf{16}, 729 (2017).

\bibitem{SunZho2009}
Y.-S. Sun and L.-Y. Zhou, \emph{Stickiness in three-dimensional volume
  preserving mappings}, Celest.~Mech.~Dyn.~Astron. \textbf{103}, 119 (2009).

\bibitem{SilBeiMan2015}
R.~M. da~Silva, M.~W. Beims, and C.~Manchein, \emph{Recurrence-time statistics
  in non-{Hamiltonian} volume-preserving maps and flows}, Phys.~Rev.~E
  \textbf{92}, 022921 (2015).

\bibitem{MeiMigSimVie2018}
J.~D. Meiss, N.~Miguel, C.~Sim\'o, and A.~Vieiro, \emph{Accelerator modes and
  anomalous diffusion in {3D} volume-preserving maps}, Nonlinearity
  \textbf{31}, 5615 (2018).

\bibitem{DomFriGreHenMehSow1986}
T.~Dombre, U.~Frisch, J.~M. Greene, M.~H\'enon, A.~Mehr, and A.~M. Soward,
  \emph{Chaotic streamlines in the {ABC} flows}, J.~Fluid Mech. \textbf{167},
  353 (1986).

\bibitem{FeiKadPir1988}
M.~Feingold, L.~P. Kadanoff, and O.~Piro, \emph{Passive scalars,
  three-dimensional volume-preserving maps, and chaos}, J.~Stat.~Phys.
  \textbf{50}, 529 (1988).

\bibitem{Arn1965}
V.~Arnold, \emph{Sur la topologie des ecoulements stationnaires des fluides
  parfaits}, C. R. Acad Sc. Paris \textbf{261}, 17 (1965).

\bibitem{Chi1970}
S.~Childress, \emph{New solutions of the kinematic dynamo problem},
  J.~Math.~Phys. \textbf{11}, 3063 (1970).

\bibitem{RamDasKriMit2014}
A.~K. Ram, B.~Dasgupta, V.~Krishnamurthy, and D.~Mitra, \emph{Anomalous
  diffusion of field lines and charged particles in
  {Arnold}-{Beltrami}-{Childress} force-free magnetic fields}, Phys.~Plasmas
  \textbf{21}, 072309 (2014).

\bibitem{CarMagPirTuv2002}
J.~H.~E. Cartwright, M.~O. Magnasco, O.~Piro, and I.~Tuval, \emph{Bailout
  embeddings and neutrally buoyant particles in three-dimensional flows},
  Phys.~Rev.~Lett. \textbf{89}, 264501 (2002).

\bibitem{DasGup2014}
S.~Das and N.~Gupte, \emph{Dynamics of impurities in a three-dimensional
  volume-preserving map}, Phys.~Rev.~E \textbf{90}, 012906 (2014).

\bibitem{DasGup2017}
S.~Das and N.~Gupte, \emph{Transport, diffusion, and energy studies in the
  {Arnold}-{Beltrami}-{Childress} map}, Phys.~Rev.~E \textbf{96}, 032210
  (2017).

\bibitem{LanRicOnkBaeKet2014}
S.~Lange, M.~Richter, F.~Onken, A.~B\"acker, and R.~Ketzmerick, \emph{Global
  structure of regular tori in a generic {4D} symplectic map}, Chaos
  \textbf{24}, 024409 (2014).

\bibitem{GreMacSta1986}
J.~M. Greene, R.~S. MacKay, and J.~Stark, \emph{Boundary circles for
  area-preserving maps}, Physica~D \textbf{21}, 267 (1986).

\bibitem{EckRue1985}
J.~P. Eckmann and D.~Ruelle, \emph{Ergodic theory of chaos and strange
  attractors}, Rev.~Mod.~Phys. \textbf{57}, 617 (1985).

\bibitem{SzeLopVia2005}
J.~D. {Szezech Jr.}, S.~R. Lopes, and R.~L. Viana, \emph{Finite-time {L}yapunov
  spectrum for chaotic orbits of non-integrable {H}amiltonian systems},
  Phys.~Lett.~A \textbf{335}, 394 (2005).

\bibitem{ManBeiRos2012}
C.~Manchein, M.~W. Beims, and J.~M. Rost, \emph{Characterizing the dynamics of
  higher dimensional nonintegrable conservative systems}, Chaos \textbf{22},
  033137 (2012).

\bibitem{Las1993}
J.~Laskar, \emph{Frequency analysis for multi-dimensional systems. {G}lobal
  dynamics and diffusion}, Physica~D \textbf{67}, 257 (1993).

\bibitem{BarBazGioScaTod1996}
R.~Bartolini, A.~Bazzani, M.~Giovannozzi, W.~Scandale, and E.~Todesco,
  \emph{{{T}une evaluation in simulations and experiments}}, Part.~Accel.
  \textbf{52}, 147 (1996).

\bibitem{SkoGotLas2016}
{\relax Ch}.~Skokos, G.~A. Gottwald, and J.~Laskar (editors) \emph{Chaos
  Detection and Predictability}, volume 915 of \emph{Lecture Notes in Physics},
  {Springer Berlin Heidelberg} (2016).

\bibitem{DulMei2009}
H.~R. Dullin and J.~D. Meiss, \emph{Quadratic volume-preserving maps: Invariant
  circles and bifurcations}, {SIAM} J.~Appl.~Dyn.~Syst. \textbf{8}, 76 (2009).

\bibitem{OnkLanKetBae2016}
F.~Onken, S.~Lange, R.~Ketzmerick, and A.~B\"acker, \emph{Bifurcations of
  families of {1D}-tori in {4D} symplectic maps}, Chaos \textbf{26}, 063124
  (2016).

\bibitem{Poi1885}
H.~Poincar\'e, \emph{Sur l'\'equilibre d'une masse fluide anim\'ee d'un
  mouvement de rotation}, Acta mathematica \textbf{7}, 259 (1885).

\bibitem{Kay1983}
R.~S. MacKay, \emph{A renormalization approach to invariant circles in
  area-preserving maps}, Physica~D \textbf{7}, 283 (1983).

\bibitem{MeiOtt1986}
J.~D. Meiss and E.~Ott, \emph{Markov tree model of transport in area-preserving
  maps}, Physica~D \textbf{20}, 387 (1986).

\bibitem{RamVar2011}
P.~Ramachandran and G.~Varoquaux, \emph{{Mayavi}: {3D} visualization of
  scientific data}, Comput.~Sci.~Eng. \textbf{13}, 40 (2011).

\end{thebibliography}

\end{document}